\documentclass[prl,twocolumn,superscriptaddress,showpacs, nofootinbib]{revtex4-2}
\usepackage{graphicx}% Include figure files
\usepackage{float}
\usepackage[utf8]{inputenc}
\usepackage{dcolumn}% Align table columns on decimal point
\usepackage{float}
\usepackage{bm}% bold math
\usepackage{mathrsfs}
\usepackage{ulem}
\usepackage[per-mode = fraction]{siunitx}
\usepackage{hyperref}
\usepackage[american]{babel}
\usepackage{booktabs}
\hypersetup{
    colorlinks=true,
    linkcolor=blue,
    filecolor=magenta,      
    urlcolor=cyan,
    citecolor=magenta,
}
\usepackage{xcolor}
\usepackage{soul}
\usepackage{amsthm,amssymb}
\usepackage{mathtools}
\usepackage{comment}
\usepackage{physics}
\providecommand{\U}[1]{\protect\rule{.1in}{.1in}}

\begin{document}
\title{Microwave Signature of the Emerging Abrikosov Lattice Above $H_{c2}$}
\author{Hang Zhou}
\affiliation{Department of Physics, College of Physical Science and Technology, Xiamen University, Xiamen 361005, China}
\author{Zhanghai Chen}
\affiliation{Department of Physics, College of Physical Science and Technology, Xiamen University, Xiamen 361005, China}
\author{A. A. Varlamov}
\affiliation{CNR-SPIN, Via del Fosso del Cavaliere, 100, 00133 Rome, Italy}
\affiliation{Department of Physics, College of Physical Science and Technology, Xiamen University, Xiamen 361005, China}
\author{Andreas Glatz}
\affiliation{Argonne National Laboratory, S.Cass Ave. 9700, Argonne, Illinois 60439, USA}
\affiliation{Department of Physics, Northern Illinois University, DeKalb, Illinois 60115, USA}
\author{Yuriy Yerin}
\affiliation{Istituto di Struttura della Materia of the National Research Council, via Salaria Km 29.3, I-00016 Monterotondo Stazione, Italy}
\date{\today}
\begin{abstract} The emergence of the Abrikosov lattice in the normal phase of type-II superconducting films when the magnetic field approaches the critical field $H_{c2}$ from above was predicted in Ref.~\cite{GVV2011}. 
In the quantum fluctuation regime \cite{GL2001} it is characterized by the formation of relatively large (with sizes of order $\xi_{\mathrm{QF}} \sim \xi_{\mathrm{BCS}}\sqrt{H_{c2}/(H-H_{c2})}$) ``long lived''  (lifetime of order $\tau_{\mathrm{QF}} \sim \hbar \Delta^{-1} H_{c2}/(H-H_{c2})$) clusters of rotating fluctuation Cooper pairs - signatures of developing Abrikosov vortices. 
We demonstrate that these fluctuation-induced vortex clusters, previously considered unobservable due to their ultrafast dynamics and weak (only logarithmically  singular) contribution to the dc-conductivity, can in fact be detected through their distinct electromagnetic signature. By analyzing the high-frequency electromagnetic response of these rotating fluctuation Cooper pairs above the second critical field in superconducting film, we predict a pronounced and measurable enhancement in the imaginary part of the ac-conductivity arising directly from quantum fluctuations. 
This enhancement is expected to occur at characteristic frequencies $\omega_{QF} \sim \hbar^{-1}\Delta(H-H_{c2})/H_{c2}$, which are well below the superconducting threshold at $2\hbar^{-1}\Delta $, where a similar increase in imaginary conductivity occurs in the superconducting phase.   
For niobium, a prototypical type II superconductor, $\omega_{QF}$ lies in the experimentally accessible microwave range, making the effect directly testable with modern microwave spectroscopy. 
\end{abstract}

\maketitle

{\it Introduction.} Gottfried Liebniz cautioned as early as 1704: ``Natura non facit saltus'' \cite{Liebniz}.  This warning in breakthrough solutions of long-standing fundamental problems such as the theory of second-order phase transitions (Landau \cite{Landau37}), the theory of superconductivity (Bardeen, Cooper, Schrieffer \cite{BCS}), etc. is usually ignored at first. The mentioned mean-field theories describe the relevant phenomena often successfully, depending on the material parameters of those or other systems. Forced to go beyond them usually requires the development of new ideas: scaling theory and the renormalization group approach in the case of extension beyond Landau theory \cite{PP79} (which breaks down for example when describing the $\lambda$-point in $^4$He \cite{HHS1976}), advancing beyond BCS theory in the framework of the Ginzburg-Landau functional or diagrammatic methods in studying the properties of low-dimensional and impure superconductors \cite{LV09}.

\begin{figure}
	\includegraphics[width=0.99\columnwidth]{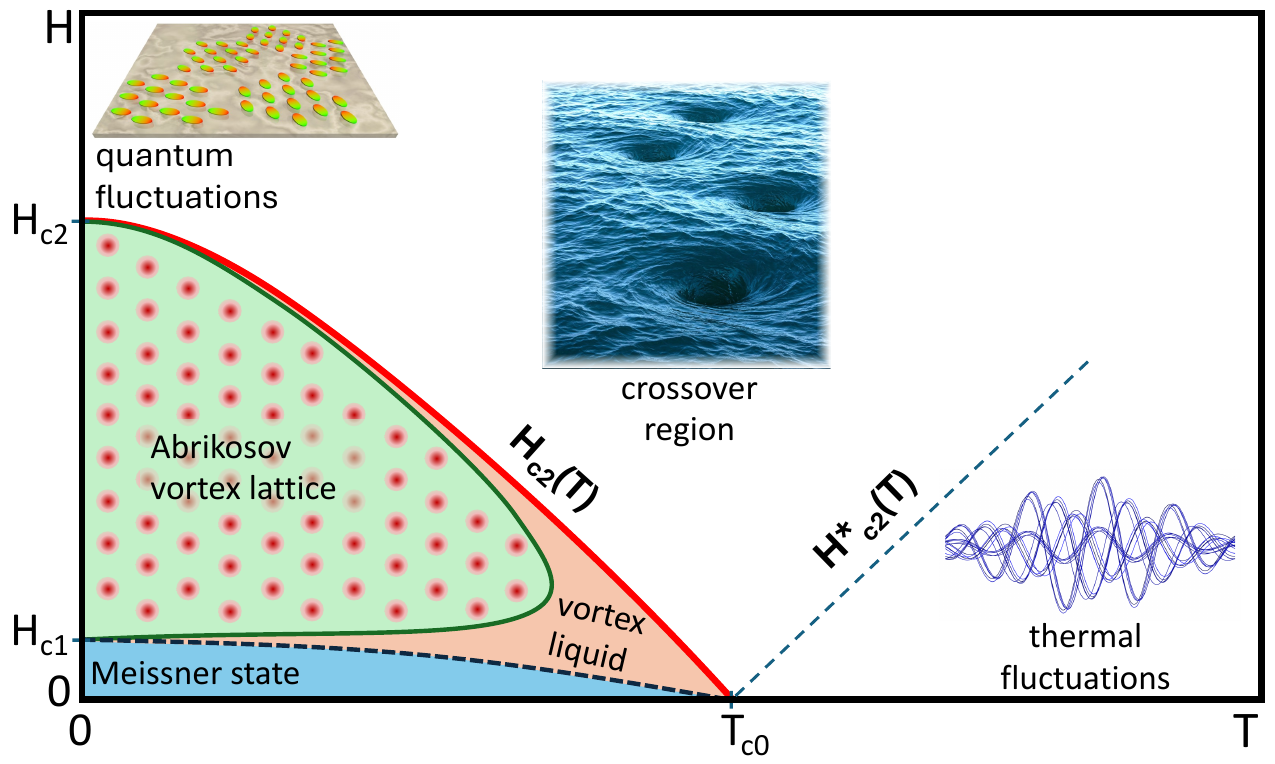}
	\caption {Schematic phase diagram of type-II superconductors, showing the domains of qualitatively different behavior of fluctuations. Here $H_{c2}(T)$ is the temperature dependent second critical field, while $H^*_{c2}(T)$ is the mirror field, where the magnetic length of the FCP equates to the Ginzburg-Landau length~\cite{GPV2020}.}
\label{phasediagram0}
\end{figure}
	
Since their extensive investigation began in the 1960s \cite{AL68,M68,T70,LV09}, thermodynamic fluctuations in superconductors have become well understood, offering a solid quantitative framework for characterizing superconducting materials via their fluctuation properties.  Namely, when temperature decreases and the system approaches its transition point to superconducting state, the electron correlations in Cooper channel result in formation of non-equilibrium Cooper pairs already in the normal state. Their concentration, lifetime, and characteristic coherence length $ \xi_{\mathrm{GL}}(T)$ strongly depend on proximity to the transition. In spite of the temporal character of these entities, they succeed in contributing to all thermodynamic and transport characteristics of metal reflecting the approaching superconducting state: growth of heat capacity, conductivity, and diamagnetic susceptibility, decrease of the electron density of states (DOS), opening of the pseudogap in quasiparticle spectrum, etc. 

In the absence of magnetic field, close to $T_{c0}$ these fluctuations can be imaged as long wavelength modes of the order parameter oscillations with $\lambda \sim \xi_{\mathrm{GL}}(T) \sim \xi_{\mathrm{BCS}}/\sqrt{(T-T_{c0})/T_{c0}}$, where $\xi_{\mathrm{BCS}}$ is the BCS coherence length at $T=0$. The lifetime of such correlations (which we will call fluctuation Cooper pairs, FCP) also grows when approaching $T_{c0}$ as $\tau_{\mathrm{GL}}=\pi \hbar/8k_B(T-T_{c0})$. This fact shows that the Cooper pairs have to become stable when the temperature decreases below the critical one. 

The picture described above remains valid moving along the line $H_{c2}(T)$ with the growth of magnetic field as long as the corresponding magnetic length $l_H=\sqrt{c\hbar/2eH}$ remains large enough. When $l_H\sim \xi_{\mathrm{GL}}(T)$, the system enters in the crossover regime, where the Ginzburg-Landau picture of the harmonic modes ceases to be adequate: under the effect of the magnetic field the waves gradually break into separate vortices (see Fig.~\ref{phasediagram0}).

The domain of phase diagram where the temperature approaches zero and the magnetic field is slightly higher than $H_{c2}(0)$ is the region of purely quantum fluctuations.  Here, the emerging fluctuating Cooper pairs, under the influence of a strong magnetic field, are coerced to rotate with the frequency $\omega_c(H) \approx 2\Delta/\hbar $, becoming prototypes of future Abrikosov vortices  ($\Delta$ is the BCS value of the gap at zero temperature) \cite{GVV2011,GVV2012}. 
However, the times of superconducting correlations near $H_{c2}(0)$ turn out to be much longer than the rotation period $\tau_\Delta=\pi \hbar/\Delta$ of the Cooper pair in the vortex: their characteristic lifetime is $\tau_{\mathrm{QF}} \sim \hbar/ (\Delta \tilde h ) \gg \tau_{\Delta}$, where $\tilde h =(H-H_{c2})/H_{c2} \ll 1$ is the reduced magnetic field, playing near $H_{c2}$ a similar role as the reduced temperature $\epsilon = (T-T_{c0})/T_{c0}$ in the closeness of $T_{c0}$. As a result, relatively large clusters (of size $\xi_{\mathrm{QF}} \sim \xi_{\mathrm{BCS}}/\sqrt{\tilde h }$), containing a large number of rotating FCPs, arise in the system (see Fig. \ref{phasediagram0} )\cite{GVV2011,GVV2012}.

\begin{figure}
\begin{center}
\includegraphics[width=0.95\columnwidth]{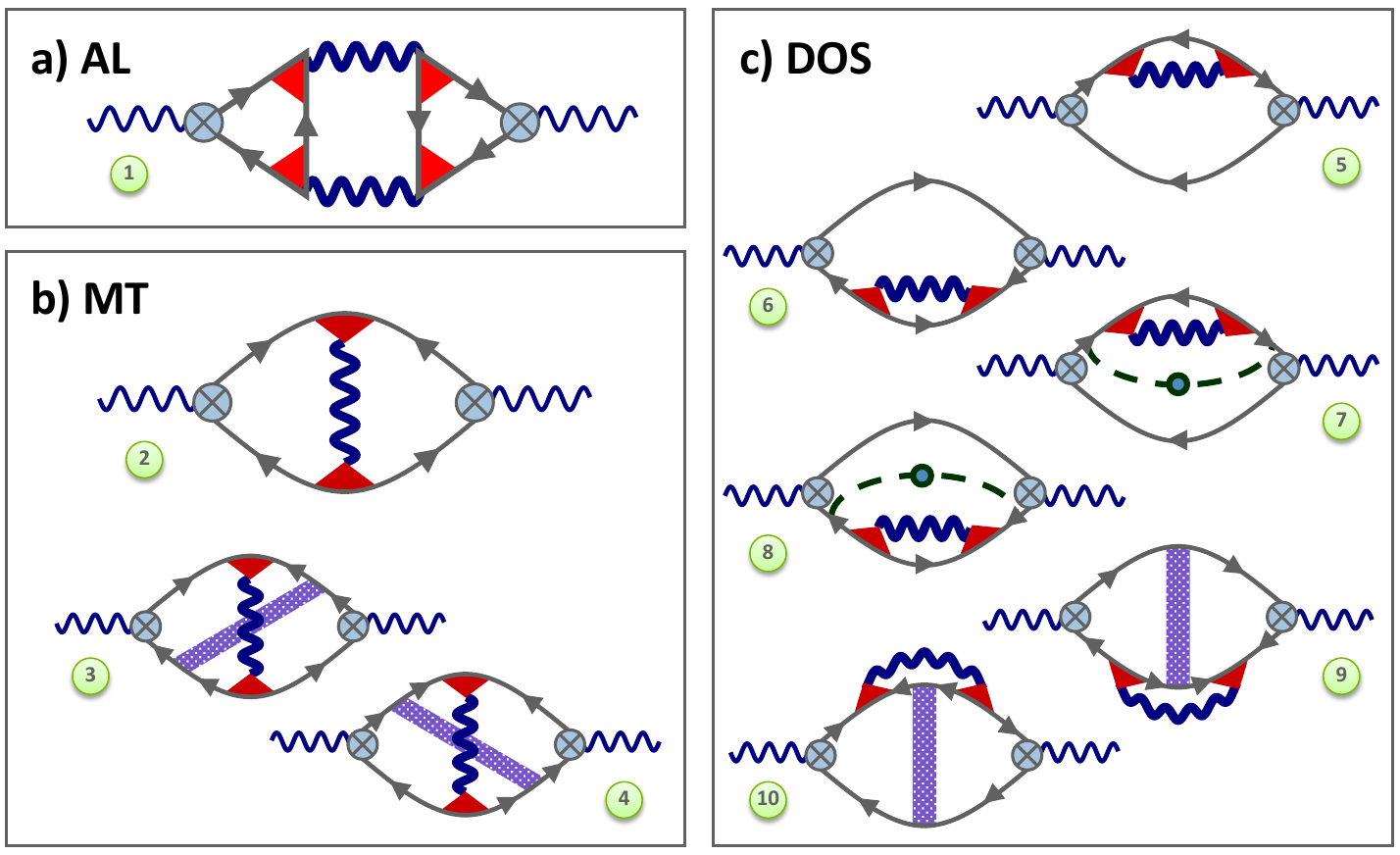}
\end{center}
\caption{(Color online) Feynman diagrams for the leading-order contributions to the electromagnetic response operator. Wavy lines correspond to fluctuation propagators, solid lines with arrows represent impurity-averaged normal state electron Green's functions, crossed circles are electric field vertices, dashed lines with a circle represent additional impurity renormalizations, and triangles and dotted rectangles are impurity ladders accounting for the electron scattering at impurities (Cooperons, see Supplementary Material \cite{SM} for additional details).}
\label{fig.conddia}
\end{figure}
In order to demonstrate the emergence of the Abrikosov lattice precursors still in the normal phase of a type II superconducting film, we study the response of such clusters of rotating FCPs to applied electromagnetic radiation. The properties of a superconductor in a weak high-frequency electromagnetic field were first investigated in a seminal paper by Abrikosov, Gorkov, and Khalatnikov \cite{AGK58}. The authors have shown that,  according to the physical meaning of the gap, the absorption of electromagnetic radiation in the absence of quasiparticle excitations ($T=0$) can occur only due to pair destruction, i.e., the real part of the ac-impedance starts to differ from zero at frequencies $\omega \geq 2\hbar^{-1}\Delta$. As for the imaginary part of conductivity, in the absence of magnetic field it emerges already at low frequencies, grows linearly with an increase in frequency in the range $\omega \lesssim 2\hbar^{-1}\Delta$ and reaches its maximum at $\omega \sim 2\hbar^{-1}\Delta$ \cite{AGK58}. Kazumi Maki reached the same conclusion considering  the vortex phase just below $H_{c2}(0)$ \cite{M66}.The imaginary part is always positive, but it grows significantly beyond the Drude value in the normal phase, which reflects the accumulation of inductive electromagnetic energy caused by screening currents \cite{MattisBardeen1958, Tinkham1996, CC91}.

In this Letter, we show that the presence of FCPs in a superconducting film above $H_{c2}(0)$ manifests itself as a characteristic growth of the value of imaginary part of {\it ac}-conductivity at frequency $\omega_{\mathrm{QF}} \sim \hbar^{-1}\Delta \tilde h$. This value is much smaller than the one for the same sample in its superconducting state [$\omega_c(H_{c2}) = 2\hbar^{-1}\Delta$], which in turn is much smaller than the peak position in the imaginary part of the Drude conductivity of the normal metal ($\omega_{\mathrm{Dr}} \sim \tau^{-1} \gg \omega_c \gg \omega_{\mathrm{QF}}$, where $\tau$ is the elastic scattering time of electrons on impurities). For example, for niobium, a reference type-II superconductor, these frequencies correspond to a range of $0.1-1$ GHz.

{\it The hierarchy of quantum fluctuations above $H_{c2}(0)$.}
Complete analysis of the effect of quantum fluctuations on the {\it dc}-conductivity in the vicinity of $H_{c2}(T)$ was performed in Ref.~\cite{GL2001,GVV2011,TSF2012}. It turned out that the hierarchy of the different contributions in the region of quantum fluctuations ($T=0$, $H-H_{c2}(0) \ll H_{c2}(0)$) differs drastically from that of thermodynamic fluctuations close to $T_{c0}$. Above it, the dominant contributions to conductivity are due to the Aslamazov–Larkin (AL) and the so-called anomalous Maki–Thompson (MT) processes (diagrams 1 and 2 in Fig.~\ref{fig.conddia}). In contrast, above $H_{c2}(0)$, in the quantum fluctuation regime, these processes do not contribute to the dc conductivity. Here it is the regular part of the 
Maki-Thompson diagram and its modifications, which take into account multiple electron scattering on impurities by means of four-leg Cooperons (diagrams 2-4 in Fig.~\ref{fig.conddia}), play a major role, giving the negative contribution, logarithmically divergent approaching $H_{c2}(0)$. In contrast, the corresponding contributions from the diagrams that account for the renormalization of the one-electron density of states (DOS) and their modifications (diagrams 5–10 in Fig.~\ref{fig.conddia}), are found to cancel out.

Addressing the contribution of quantum fluctuations to the ac-conductivity, we will show below that the leading role returns to AL and anomalous MT contributions.

{\it AC-conductivity above $H_{c2}(0)$.} 
The account for the effect of a finite frequency electromagnetic field on the fluctuation conductivity can be carried out within the same formalism of the Matsubara technique as in the {\it dc}-limit \cite{GL2001,GVV2011}. 
However, when performing the analytical continuation of the electromagnetic response operator (the diagrams shown in Fig.~\ref{fig.conddia}), one cannot restrict oneself to the limit of infinitesimal frequencies, which makes the problem considerably more cumbersome and sophisticated.

The details of calculus one can find in the Supplementary material \cite{SM}. The corresponding results for all quantum fluctuation corrections to the ac-conductivity are presented in Table I. One can see that six DOS type diagrams (diagrams 5-10 in Fig. \ref{fig.conddia}) exactly annihilate each other (see last two rows of Table I), while the three MT type diagrams contribute in equal parts to the logarithmic singularity of the dc conductivity.  The most singular contribution of fluctuations, which manifests itself in the imaginary part of the conductivity as $\sim \omega/\Delta \tilde{h}$ and which is the focus of this work, arises from AL and anomalous MT processes (see the first two rows of Table I).

The total contribution of quantum fluctuations to conductivity in the limit of low temperatures and frequencies can be outlined as follows:
\begin{widetext}
\begin{equation}
  \sigma _{xx}^{\mathrm{FL}} (t,\tilde{h},\omega \ll \Delta \tilde{h} )= \frac{e^2}{\pi ^2\hbar}\left[-6\ln\frac{1}{\tilde{h}}+\frac{2}{9}\left(\frac{\gamma_{E}t}{\tilde{h}}\right)^2+\frac{5}{3}\frac{i\omega}{\Delta \tilde{h}}-\frac{7}{9}\left(\frac{\omega}{\Delta \tilde{h}}\right)^2\right].
\label{tot}
\end{equation}
%\end{widetext}
%\begin{widetext}
\begin{center}
\begin{table}[tbh]
    \centering
    \begin{tabular}{|c|c|c|}
        \hline
       Contribution &  Complete expression ($t\ll \tilde{h}\ll1;\omega \ll \Delta $) & Low frequency $(\omega\ll\Delta\tilde{h})$ asymptotic \\
       \hline
       $\tilde \sigma_{xx}^{\mathrm{AL}}$  & $\frac{16}{9}\widetilde{h}\ln\frac{1}{\widetilde{h}}+\frac{8}{9}\left(\frac{\gamma_{E}t}{\widetilde{h}}\right)^{2}-\frac{4}{3}\left[1+\frac{\Delta\widetilde{h}}{i\omega}\ln\left(1-\frac{i\omega}{\Delta\widetilde{h}}\right)\right]$ & $\frac{16}{9}\tilde{h}\ln\frac{1}{\tilde{h}}\!+\!\frac{8}{9}\left(\frac{\gamma_{E}t}{\tilde{h}}\right)^{2}\!+\!\frac{2i\omega}{3\Delta \tilde{h}}\!-\!\frac{4}{9}\left(\frac{\omega}{\Delta\tilde{h}}\right)^{2}$ \\
        \hline 
        $\tilde \sigma_{xx}^{\mathrm{MT}(an)}$ & $\frac{4}{3}\left(\frac{\gamma_{E}t}{\widetilde{h}}\right)^{2}+2\left[\frac{\Delta\widetilde{h}}{i\omega}\left(1-\frac{i\omega}{\Delta\widetilde{h}}\right)\ln\left(1-\frac{i\omega}{\Delta\widetilde{h}}\right)+1\right]
        $ & $\frac{4}{3}\left(\frac{\gamma_{E}t}{\widetilde{h}}\right)^{2}+\frac{i\omega}{\Delta\widetilde{h}}-\frac{1}{3}\left(\frac{\omega}{\Delta\widetilde{h}}\right)^{2}$ \\
        \hline
        $\tilde \sigma_{xx}^{\mathrm{MT}(reg)}$  & $-2\left[\ln\frac{1}{\tilde{h}}\!+\!\frac{1}{3}\left(\frac{\gamma_{E}t}{\tilde{h}}\right)^2\right]\left[\frac{\Delta^{2}}{2\omega^{2}}\ln\left(1-\frac{2i\omega}{\Delta}\right)-\frac{\Delta}{i\omega}\right]$ & $-2\left[\ln\frac{1}{\tilde{h}}\!+\!\frac{1}{3}\left(\frac{\gamma_{E}t}{\tilde{h}}\right)^2\right]\left(1\!+\!\frac{4i\omega}{3\Delta}-\frac{2\omega^2}{\Delta^2}\right)$ \\
         % \hline
          $\tilde\sigma_{xx}^{\mathrm{MT}(3-4)}$  & $2\left[\ln\frac{1}{\widetilde{h}}+\frac{1}{3}\left(\frac{\gamma_{E}t}{\widetilde{h}}\right)^{2}\right] \left[\left(\frac{\Delta}{i\omega}\right)^{3}\ln\frac{1-\frac{2i\omega}{\Delta}}{\left(1-\frac{i\omega}{\Delta}\right)^{2}}+\frac{\Delta}{i\omega}\right] $ & $-4\left[\ln\frac{1}{\tilde{h}}\!+\!\frac{1}{3}\left(\frac{\gamma_{E}t}{\tilde{h}}\right)^2\right]\left(1+\frac{{7i\omega}}{{4{\Delta}}}-\frac{3\omega^2}{\Delta^2}\right)$ \\
       \hline
        $\tilde \sigma_{xx}^{\mathrm{DOS}(5-8)}$ & $-2\left[\ln\frac{1}{\tilde{h}}\!+\!\frac{1}{3}\left(\frac{\gamma_{E}t}{\tilde{h}}\right)^2+\ln\left(1-\frac{i\omega}{\Delta\widetilde{h}}\right)\right]$ & $-2\left[\ln\frac{1}{\tilde{h}}\!+\!\frac{1}{3}\left(\frac{\gamma_{E}t}{\tilde{h}}\right)^2\right]+\frac{2i\omega}{\Delta \tilde{h}}-\left(\frac{\omega}{\Delta \tilde h}\right)^2$ \\
       % \hline
        $\tilde\sigma_{xx}^{\mathrm{DOS}(9-10)}$  & $ 2\left[\ln\frac{1}{\widetilde{h}}+\frac{1}{3}\left(\frac{\gamma_{E}t}{\tilde{h}}\right)^2+\ln\left(1-\frac{i\omega}{\Delta\widetilde{h}}\right)\right]$ & $2\left[\ln\frac{1}{\tilde{h}}\!+\!\frac{1}{3}\left(\frac{\gamma_{E}t}{\tilde{h}}\right)^2\right]-\frac{2i\omega}{\Delta\widetilde{h}}+\left(\frac{\omega}{\Delta \tilde h}\right)^2$\\
        \hline
    \end{tabular}
\parbox{\textwidth}{\caption{Different fluctuation contributions as the function of magnetic field, temperature, and frequency in the regime of quantum fluctuations $(T\ll\Delta\tilde{h})$. Here $\tilde \sigma_{xx}^{(i)}=\frac{\pi^2 \hbar}{e^2} \sigma_{xx}^{(i)}$, $t=T/T_{\mathrm{c0}}$, and $\gamma_E=0,577...$ is the Euler-Mascheroni constant.}}
\label{results} 
\end{table}
\end{center}
\end{widetext}

%\end{widetext}
%\begin{widetext}
\begin{figure*}
\includegraphics[width=0.95\textwidth]{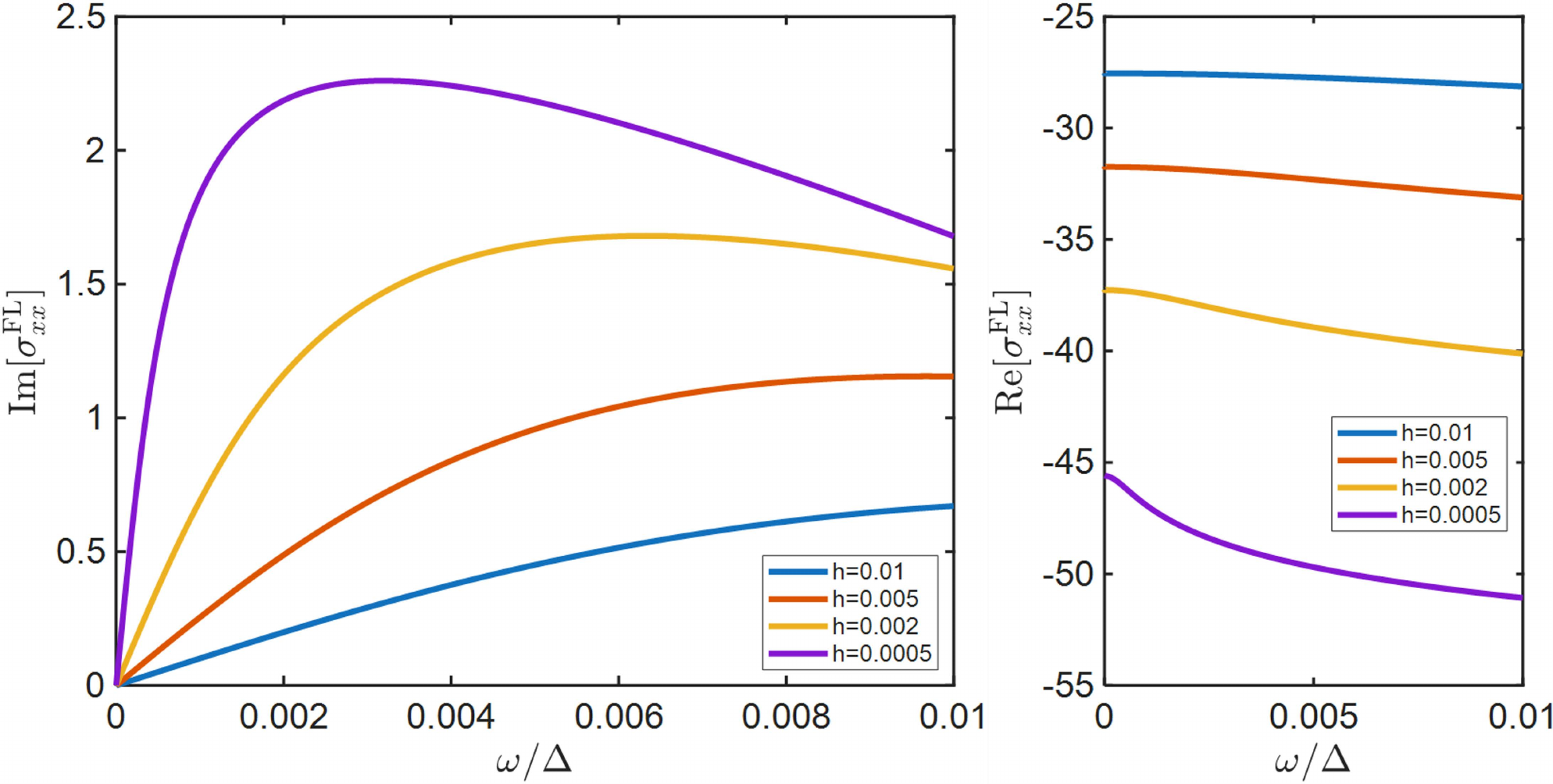}
\caption{Imaginary (left) and real (right) parts of total fluctuation conductivity $\sigma$ in units of $e^2/\hbar$ as a function of frequency $\omega/\Delta$ for different magnetic fields $\tilde{h}$ .}	
\label{REIMtot} 
\end{figure*}

In Fig. 3, the frequency dependence of the imaginary and real parts of the fluctuation conductivity is shown. One can see that in the regime of quantum fluctuations, a pronounced increase in the value of $\Im \sigma _{xx}^{\mathrm{FL}}(\omega)$ indeed appears at $\omega \sim \hbar^{-1}\Delta \tilde{h} $, shifting progressively toward zero frequency as the magnetic field approaches the transition point ($H \rightarrow H_{c2} (0)$).

For niobium, where $H_{c2}(0) \approx 2 \cdot 10^3$ ${{\O{}}}$e and $\Delta \approx 2.32 $ meV (corresponding to 560 GHz), this characteristic feature of $\Im \sigma^{tot}_{xx}(\omega)$ occurs at approximately $\omega \approx 0.5,\text{GHz}$ for $H-H_{c2}=20 $ ${{\O{}}}$e and $\omega \approx  0.1$ GHz for $H-H_{c2}=2 $ ${{\O{}}}$e. These frequencies are about three orders of magnitude lower, making them much more accessible for experimental observation \cite{footnote}.

{\it Qualitative insights.} We start with the estimation of the ratio of the concentration $N_{\mathrm{FCP}}$ of FCPs to their mass $m^*$ in the region of quantum fluctuations. In the proximity of $T_{c0}$ this is trivial, one can just integrate the Bose-Einstein distribution with the Ginzburg-Landau energy for FCP (see \cite{VGG2018}), and then relate it to the trace of fluctuation propagator $L(q,|\Omega_{k}|)$:
\begin{align}
\frac{N_{\mathrm{FCP}}(\epsilon)}{m^{*}}=-\frac{\nu_0}{m^{*}T_{c0}}\mathrm{Tr}\left\{ L(q,|\Omega_{k}|)\right\} =\frac{T_{c0}}{2\pi\hbar^{2}}\ln\frac{1}{\epsilon},
%\label{Nm}
\nonumber
\end{align}
where $\epsilon=\frac{T-T_{c0}}{T_{c0}}$, $\nu_0$ is the electron density of states, and $\mathrm{Tr}\left\{ ...\right\} =T\sum_{k}\int\frac{d^{2}q} {\left(2\pi\hbar\right)^{2}}\left\{ ...\right\}$ in the frequency-momentum representation. 

One can apply the analogous procedure to the case of quantum fluctuations, where in Landau representation $\mathrm{Tr}\left\{ ...\right\} =T\sum_{k}\left(\frac{eH_{c2}}{\pi}\right)\sum_n \left\{ ...\right\}$. Close to $H_{c2}(0)$ one can restrict consideration by the Lowest Landau Level approximation ($n=0$), when the fluctuation propagator acquires the simple form:
\begin{equation}
L_{0}\left( \Omega_k\right) =-\nu_{0}^{-1}\frac{1}{\widetilde{h}+|\Omega_k|
/\Delta}.
\nonumber
\end{equation}
and perform summation over bosonic frequencies $\Omega_k$
\begin{align}
&\frac{N_{\mathrm{FCP}}(\tilde h,t)}{m^*}=-\frac{\nu_0}{m^{*}\Delta}\mathrm{Tr}\left\{ L_n(|\Omega_{k}|)\right\} \nonumber  \\&\approx\left(\frac{2eH_{c2}}{\pi^2\hbar^{2}}\right)\!\sum_{k\!=\!-\frac{\Delta}{2\pi T}}^{\frac{\Delta}{2\pi T}}\frac{1}{|k|+\frac{\tilde{h}\Delta}{2\pi T}}\!=\!\frac{2\Delta}{\pi\hbar^{2}}\left[\ln\frac{1}{\tilde{h}}\!+\!\frac{1}{3}\left(\frac{\gamma_{E}t}{\tilde{h}}\right)^2\right].
\label{NFCP}
\end{align}

Having Eq. (\ref{NFCP}) at hand, we can qualitatively reproduce the various contributions listed in Table I.

For example, the AL contribution—which reflects the direct effect of FCPs on the conductivity—can be obtained from the standard Drude expression for the longitudinal conductivity in a magnetic field \cite{Abrikosov}, using the  just found ratio $N_{\mathrm{FCP}}/m^*$.  The charge must, of course, be taken as $2e$. As for impurity scattering, it presents no obstacle for FCPs. The role of scattering time is played here by the lifetime $\tau_{\mathrm{QF}}$ — scattering or decaying are the same thing. Bearing in mind that $\omega_c \tau_{QF}=2\Delta \tau_{QF}/\hbar \gg 1$ one finds
\begin{eqnarray}
%&&\sigma^{(Dr)}_{xx} = \frac{N_ee^2}{m_e} \frac{\tau}{1\!+\!\left(\omega^{(e)}_c \tau\right)^2}\rightarrow \nonumber \\
\sigma^{(AL)}_{xx}(t=0, \tilde{h}, \omega=0)&=&\frac{N_{\mathrm{FCP}}(2e)^2}{m^*} \frac{\tau_{QF}}{1\!+\!\left(2\Delta \tau_{QF}/\hbar\right)^2} \nonumber \\
&\sim& \frac{e^2}{\pi \hbar}\tilde{h}\ln\frac{1}{\tilde{h}}, 
\nonumber
\end{eqnarray}
 what with the accuracy to the numerical factor reproduces the corresponding expression from Table I. This contribution is not singular and disappears as $\tilde{h}\rightarrow 0$ in the regime of quantum fluctuations.

The anomalous MT process arises from the pairing of two electrons moving in opposite directions along a self-intersecting trajectory \cite{LV09,VGG2018}. This process contributes to conductivity only when the corresponding traversal time significantly exceeds the lifetime of the FCP, while being still limited by the phase-breaking time $\tau_\phi$. In the present case, the application of a magnetic field effectively sets 
$\tau_\phi$ equal to the rotation period $\tau_\Delta \sim \hbar \Delta^{-1} \ll \tau_{QF}$ making the realization of the anomalous MT process impossible in the quantum fluctuation regime.

Regarding the remaining regular MT, DOS and related to them contributions (diagrams 2-10), they can be associated to the decrease of the one-electron conductivity as a result of the involvement of some part of electrons in the fluctuation Cooper pairings. The corresponding changes can also be estimated based on the Drude formula, which in this case is read as ($\omega_c \tau \ll 1$)
\begin{eqnarray}
\sigma^{(2-10)}_{xx} (t\ll \tilde{h}, \omega=0)\! =\!-\frac{\delta N_ee^2}{m_e} \frac{\tau}{1\!+\!\omega^2_c \tau^2} \nonumber \\
\approx\!-\frac{2N_{\mathrm{FCP}}e^2 \tau}{m^*}\left(\frac{m^*}{m_e}\right) .
\nonumber
\end{eqnarray}
The ratio $N_{\mathrm{FCP}}/m^*$ is given by Eq. (\ref{NFCP}), while $m^*/m_e$ can be estimated based on the Ginzburg-Landau relation between mass $m^*$  and coherence length $\xi^2=\xi_{\mathrm{BCS}} l$:
\begin{equation}
    4\alpha T_c  m^* \xi^2 = \hbar^2.
\nonumber\end{equation}
Here $l=v_F \tau$ is the mean free path, $\alpha =\frac{4\pi^2}{7\zeta(3)}\frac{T_c}{\epsilon_F}$ is the coefficient of the GL functional in the case of BCS normalization of the order parameter, and $\xi_{\mathrm{BCS}} =\frac{\hbar v_F}{\pi \Delta}$ \cite{LV09}. 
In result, one finds that
\begin{equation}
   \sigma^{(2-10)}_{xx}(t \ll \tilde{h},\omega=0) \approx \!-\frac{e^{2}}{4\hbar}\left[\ln\frac{1}{\tilde{h}}\!+\!\frac{1}{3}\left(\frac{\gamma_{E}t}{\tilde{h}}\right)^2\right],
\label{210} 
\end{equation}
what with the accuracy of the numerical factor of the order of 2 is in complete agreement with our microscopic findings (see Table I).

As the temperature deviates from zero, the number of FCP increases due to the thermal fluctuations (see Eq. (\ref{NFCP})) which manifests itself in Eq. (\ref{210}). Yet, besides this,  FCPs can now change their state due to the interaction with the thermal bath, i.e., their  hopping along the applied electric field to an adjacent rotation trajectory  becomes possible, which means that FCPs can now participate in the longitudinal charge transfer by means of AL and anomalous MT processes. 

As an example, one of these processes can be mapped to the paraconductivity of granular superconductors~\cite{LVV08} at temperatures above $T_{\mathrm{c0}}$, where the tunneling of FCPs is determined by the conditional probability of two-electron hopping and is proportional to $W_{\Gamma} = \Gamma^{2}\tau_{\mathrm{GL}}$ (where $\Gamma $ is the rate of intergrain electron tunneling). 
Returning to the situation of FCPs above $H_{\mathrm{c2}}(0)$, one can identify the tunneling rate by the temperature $T$, while $\tau_{\mathrm{GL}}$ corresponds to $\tau_{\mathrm{QF}}$: the related conditional probability is $W_{T} = T^{2}\tau_{\mathrm{QF}}$. Estimating the direct contribution of FCPs to conductivity by replacing $\tau_{GL} \rightarrow \tau_{QF}$ in the classic AL formula gives $ \left(e^2/\hbar\right) \tau_{QF}$ and then multiplying it by $W_{T}$ one recognizes the origin of the term  $\sim \left(e^2/\hbar\right)\left(t/\tilde{h}\right)^2$ 
in the first line of Table I.

{\it Summary.} We have shown that approaching the upper critical field  $H_{c2}(0)$, quantum fluctuations  
produce a distinct enhancement in the imaginary part of the ac-conductivity  at frequencies \(\omega_{QF} \sim \hbar^{-1}\Delta \tilde{h}\), which are orders of 
magnitude lower than the characteristic scales of both the superconducting  state (\(\omega_{c} \sim 2\Delta/\hbar\)) and the normal Drude response. Being small compared to the Drude contribution to $\Im \sigma_n$ by the Ginzburg–Levanyuk number $Gi_{(2)} \sim 1/(p_F^2 l d)$ \cite{LV09}, the corresponding fluctuation contribution increases as $H_{c2}/(H-H_{c2})$ approaching the upper critical field, eventually matching at the edge of the critical region the Maki’s result \cite{M66} obtained in the vortex phase. 

The emergence of vortex precursors above $H_{c2}$ can be detected as a resonant frequency shift in a microwave cavity or coplanar resonator, arising from the fluctuation-induced change of kinetic inductance. Although the fluctuation correction to the imaginary part of the conductivity is smaller than the Drude background, it exhibits a pronounced field dependence $\sim 1 / \tilde{h}$ that makes it experimentally distinguishable. For niobium, this corresponds to shifts at $\omega \sim 0.1{-}1~\mathrm{GHz}$, well within the reach of modern microwave spectroscopy. By contrast, dc-transport would be far less sensitive since in that case the fluctuation correction depends only logarithmically on the field.

{\it Acknowledgments.} We acknowledge Todor Mishonov, Rufus Boyack, Sergey Budko for helpful critical discussions and comments.
AAV acknowledges COST action CA21144 SUPERQUMAP for partial financial support of this work. Y. Y. acknowledges the funding received from HPC National Center for HPC, Big Data and Quantum Computing - HPC (Centro Nazionale 01 – CN0000013). AG was supported by the U.S. Department of Energy, Office of Science, Basic Energy Sciences, Materials Sciences and Engineering Division.

\clearpage
\widetext
\onecolumngrid

\begin{center}
{\Large \bf Supplemental Material for}

{\large\it 
Microwave Signature of the Emerging Abrikosov Lattice Above $H_{c2}$}
\end{center}

\setcounter{figure}{0}
\setcounter{table}{0}
\renewcommand{\thefigure}{S\arabic{figure}}
\renewcommand{\thetable}{S\arabic{table}}

\section{Basic elements of microscopic description of SF in magnetic field}\label{sec.microscopic}

\subsection{General relations}

Let us begin by recalling the basic ideas of the microscopic description of fluctuations in the normal phase of a superconductor \cite{SM-LV09}. For this purpose, one can employ the formalism of the Matsubara diagrammatic technique \cite{AGD}. 
In the BCS theory, the electron--electron attraction leads to the reconstruction of the ground state of the electron system of a normal metal upon approaching the critical temperature from above ($T \rightarrow T_{\mathrm{c0}} + 0$). Formally, this fact is manifest by the appearance of a pole in the two-particle Green's function, or what is more convenient for our purposes, in the vertex part of the electron--electron interaction in the Cooper channel, $L(\mathbf{q}, \Omega_{k})$, which is called the \textit{fluctuation propagator} and where $\mathbf{q}$ is the momentum and $\Omega_{k} = 2\pi kT$ is the bosonic Matsubara frequency. 

\begin{figure}[tbh]
	\includegraphics[width=.9\columnwidth]{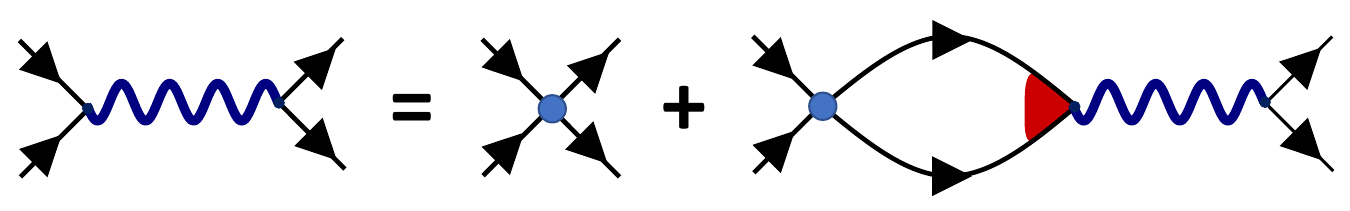}
	\caption{(Color online) The Dyson equation for the fluctuation propagator (wavy line) in the ladder approximation. Solid lines represent one-electron Green's functions, circles represent the electron--electron interaction, and the triangle corresponds to the Cooperon (see Fig.~\ref{fig.cooperon}).}
	\label{fig:S1}
\end{figure}

The Dyson equation for $L(\mathbf{q}, \Omega_{k})$, accounting for the electron--electron attraction in the ladder approximation, is represented graphically in Fig.~\ref{fig:S1}, where the wavy line corresponds to the fluctuation propagator, while the polarization operator is defined as a loop of two single-particle Green's functions in the particle--particle channel

We assume for the temperature $T \ll \min \left\{ \tau^{-1}, \omega_{\mathrm{D}}\right\}$ in order to stay both in the diffusive regime of electron scattering and in the framework of the BCS model ($\tau$ is the electron elastic scattering time at impurities). The magnitude of the magnetic field is limited by two conditions: (i) remain below the regime of Shubnikov-de Haas oscillations, $\omega_{c}\tau \lesssim 1\Longleftrightarrow H \lesssim \left( T_{\mathrm{c0}}\tau \right)^{-1}H_{\mathrm{c2}}(0)$, and (ii) stay below the Clogston limit, $H \lesssim \left( \varepsilon_{\mathrm{F}}\tau\right) H_{\mathrm{c2}}(0)$, i.e., $H/H_{\mathrm{c2}}(0) \ll \min \left\{ \left( T_{\mathrm{c0}}\tau \right)^{-1}, \varepsilon_{\mathrm{F}}\tau \right\}$.

\begin{figure}[tbh]
	\includegraphics[width=.9\columnwidth]{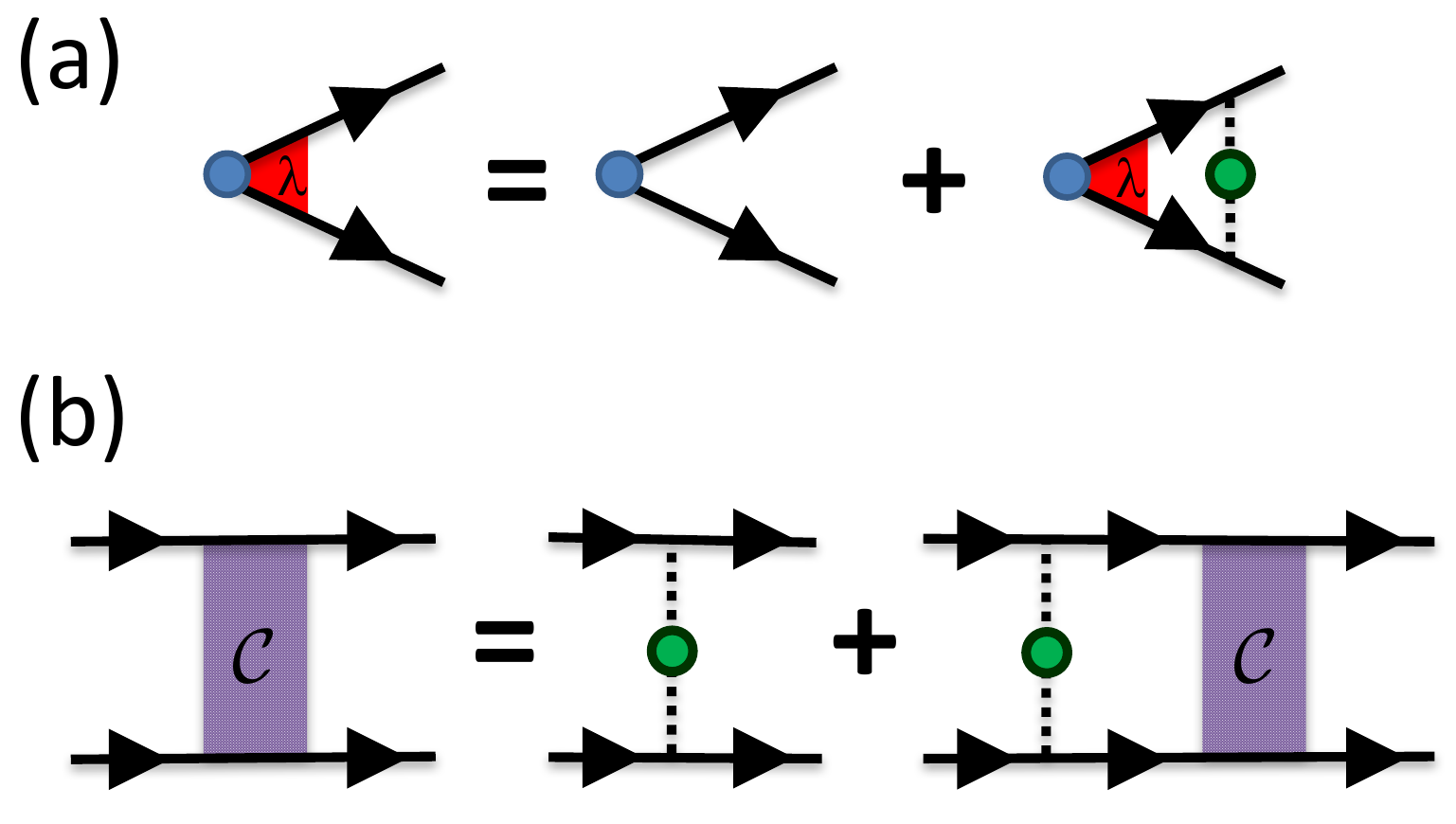}
	\caption{(Color online) (a) Dyson equation for Cooperon, i.e. the vertex that accounts for the result of averaging over elastic impurity scattering of electrons in the ladder approximation. Solid lines correspond to bare one-electron Green's functions. The dashed line is associated with an impurity correlator, $\left\langle U^{2}\right\rangle = 1/\left( 2\pi \nu_{0}\tau \right)$. (b) Analogous Dyson equation for the four-leg Cooperon in the ladder approximation.}
	\label{fig.cooperon}
\end{figure}

In addition to the appearance of the imaginary part of the self-energy in the one-particle Green's function, the effect of coherent electron scattering on impurities results in the renormalization of the vertex part in the particle--particle channel. It is determined by the Dyson equation in ladder approximation (see Fig.~\ref{fig.cooperon}).

The details of the derivations can be found in Ref.~\cite{SM-LV09}; here we only present the results necessary for further discussions. The Cooperon shown in Fig.~\ref{fig.cooperon} has the following form in Landau representation: 
\begin{equation}
	\lambda_{n}(\varepsilon_{1}, \varepsilon_{2}) = \frac{\tau^{-1}\theta (-\varepsilon_{1}\varepsilon_{2})}{|\varepsilon_{1} - \varepsilon_{2}| + \omega_{c}(n + 1/2) + \tau_{\varphi}^{-1}}\,,
	\label{lambda}
\end{equation}
where $n$ is the quantum number of the Landau state of Cooper pairs, $\theta(x)$ is the Heaviside step-function, $\varepsilon_{1}$ and $\varepsilon_{2}$ are the fermionic frequencies, and $\tau_{\varphi}$ is the phase-breaking time of electron scattering. In the process of impurity averaging, one also encounters the corresponding four-leg vertex, which differs from Eq.~(\ref{lambda}) only by the factor $ 1/({2\pi \nu_{0}\tau )}$: 
\begin{equation}
	C_{n}(\varepsilon_{1}, \varepsilon_{2}) = \frac{1}{2\pi \nu_{0}\tau }\frac{\tau^{-1}\theta(-\varepsilon_{1}\varepsilon _{2})} {|\varepsilon_{1} - \varepsilon _{2}| + \omega_{c}(n + 1/2) + \tau_{\varphi}^{-1}}\,.
	\label{cn}
\end{equation}
Finally, the expression for the fluctuation propagator in this representation takes the form: 
\begin{equation}
	 L_{m}^{-1}(\Omega_{k}) = \label{propagator} 
	 -\nu_{0}\left[ \ln\frac{T}{T_{\mathrm{c0}}} + \psi \left( \frac{1}{2} + \frac{|\Omega_{k}| + \omega_{c}(m + \frac{1}{2})}{4\pi T}\right) - \psi \left( \frac{1}{2}\right) \right]\,.  %\notag
\end{equation}
Here $\Omega_{k} = 2\pi Tk$ and $\varepsilon_{n} = (2n + 1)\pi T$ are  bosonic and fermionic Matsubara frequencies \footnote{In the following we use units with $\hbar = k_{\mathrm{B}} = c = 1$.}.

An important characteristic property of Eqs.~(\ref{lambda})--(\ref{propagator}) is that they are valid in a large region of the phase diagram of a superconductor above the line $H_{\mathrm{c2}}(T)$ for magnetic fields $H/H_{\mathrm{c2}}(0) \ll \min \left\{ \left( T_{\mathrm{c0}}\tau \right)^{-1}, \varepsilon_{\mathrm{F}}\tau \right\}$, temperatures $T \ll \min \{\tau^{-1}, \omega_{\mathrm{D}}\}$, frequencies $|\Omega_{k}| \ll \tau ^{-1}$, and Landau levels with $m \ll \left( T_{\mathrm{c0}}\tau \right)^{-1}$.

In the following, it is convenient to use the dimensionless temperature and magnetic field 
\begin{equation}
	t = \frac{T}{T_{\mathrm{c0}}}\,, \quad h = \frac{H}{\widetilde{H}_{\mathrm{c2}}(0)}.
\end{equation}
Since it is more convenient, the latter is normalized by the value of the second critical field obtained by linear extrapolation of its temperature dependence near $T_{\mathrm{c0}}$: 
\begin{equation}
	\widetilde{H}_{\mathrm{c2}}(0) = \frac{\Phi_{0}}{2\pi \xi^{2}}\,,
\end{equation}%
where $\Phi_{0} = \pi /e$ is the magnetic flux quantum. The magnetic field $\widetilde{H}_{\mathrm{c2}}(0)$ is $8\gamma_{\mathrm{E}}/\pi^{2} = 1.45$ times larger than the true second critical field $H_{\mathrm{c2}}(0)$:
\begin{equation}
	h = \frac{H}{\widetilde{H}_{\mathrm{c2}}(0)} = \frac{\pi^{2}}{8\gamma_{\mathrm{E}}}\frac{H}{H_{\mathrm{c2}}(0)} = 0.69\frac{H}{H_{\mathrm{c2}}(0)}\,.
\end{equation}

The further unification of the dimensionless units leads to $\frac{\omega_{c}}{4\pi T}=\frac{4}{\pi^{2}}\left(  \frac{h}{t}\right)
$, $M=1/T\tau$, and $\frac{\mathcal{D}}{4T}\left(  \frac{H}{\Phi_{0}}\right)
=\frac{1}{\pi^{2}}\left(  \frac{h}{t}\right)  $.

\subsection{Propagator at fields close to the line $H_{\mathrm{c2}}\left( T\right) $}

The line separating normal and superconducting phases $H_{\mathrm{c2}%
}\left( T\right) $ (in our dimensionless units the line of critical fields $%
h_{\mathrm{c2}}\left( t\right) )$ is determined by the requirement that the
propagator Eq. (\ref{propagator}) has a pole when $\Omega_{k}=0$ and $m=0$:%
\begin{equation}
\label{pole}
\ln t+\psi\left( \frac{1}{2}+\frac{2}{\pi^{2}}\frac{h_{\mathrm{c2}}\left(
t\right) }{t}\right) -\psi\left( \frac{1}{2}\right) =0.
\end{equation}

At low temperatures $T\ll T_{\mathrm{c0}},$ close to the point $T=0$ and $%
H=H_{\mathrm{c2}}(0)$, the critical field is $h_{\mathrm{c2}}\left( t\right)
=2\xi^{2}H_{\mathrm{c2}}(0)/e\sim1$. Then one can substitute the $\psi $%
-function by its asymptotic expression Eq. (\ref{psi}) and take into account that $\psi\left( 1/2\right) =-\ln4\gamma_{E}$ ($\gamma_{E}=1.781..$ is the Euler's constant) which results in
\begin{equation}
h_{\mathrm{c2}}\left( t\rightarrow0\right) =\frac{\pi^{2}}{8\gamma_{E}}.
\label{hc2}
\end{equation}

In order to find the paraconducting contribution to FC above the curve $H_{%
\mathrm{c2}}\left( T\right) $ in Fig.~\ref{phasediagram0}, let us rewrite Eq. (\ref{pole}) in terms of the reduced field
\begin{equation}
\widetilde{h}\left( t\right) =\frac{h-h_{\mathrm{c2}}\left( t\right) }{h_{%
\mathrm{c2}}\left( t\right) }\ll 1.
\end{equation}

Below we will see that the Cooper pair contribution to FC, which is singular
in $\widetilde{h}^{-1}$, originates in Eq. (\ref{propagator} ) only from the term with $m=0$, i.e. we can restrict ourselves to the Lowest Landau Level
(LLL) approximation. 
The effect of fluctuations on the conductivity at zero temperature is reduced to
the renormalization of the one-electron diffusion coefficient. FCPs in the
quantum region occupy the LLL, but all dynamic fluctuations
in the frequency interval from $0$ to $\Delta_{\mathrm{BCS}}$ have to be
taken into account. The corresponding fluctuation propagator at zero
temperature close to $H_{\mathrm{c2}}\left( 0\right) $ has the form (see Eq. (\ref{propagator}))
\begin{equation}
L_{0}\left( \Omega_k\right) =-\nu_{0}^{-1}\frac{1}{\widetilde{h}+|\Omega_k|
/\Delta_{\mathrm{BCS}}}.
\end{equation}

\subsection{Useful relations}The critical field is determined by the pole of propagator at zero temperature: 
\begin{equation}
\ln\left(\frac{\omega_{c}(H_{c2})}{8\pi T_{c0}}\right)-\psi\left(\frac{1}{2}\right)=0,
\end{equation}
\begin{equation}
\frac{\omega_{c}(H_{c2})}{8\pi T_{c0}}=\exp(\psi\left(\frac{1}{2}\right))=\exp(-\ln\left(4\gamma_{E}\right))=\frac{1}{4\gamma_{E}},
\end{equation}
\begin{equation}
\omega_{c}(H_{c2})=\frac{2\pi T_{c0}}{\gamma_{E}},
\end{equation}
\begin{equation}
\Delta_{\mathrm{BCS}}(0)=\frac{\pi T_{c0}}{\gamma_{E}}, 
\label{Delta_Tc}
\end{equation}
Hence:
\begin{equation}
\omega_{c}(H_{c2})=2\Delta_{\mathrm{BCS}}(0), 
\end{equation}
where here and hereafter, we put $\Delta_{\mathrm{BCS}}(0)=\Delta$.

Also, we provide the definition of the polygamma function
\begin{equation}
\psi^{(N)}(x)=(-1)^{N+1}N!\sum_{n=0}^{\infty}\frac{1}{(n+x)^{N+1}}, 
\label{psi_def}
\end{equation}
and asymptotic expansions at large argument $x\gg 1$ of $\psi(x)$ and $\psi '\left( x \right)$
\begin{equation}
\label{psi_half}
\psi \left( {\frac{1}{2} + x} \right) = \ln x + \frac{1}{{24{x^2}}} + O\left( {\frac{1}{{{x^3}}}} \right),
\end{equation}
\begin{equation}
\label{psi1_half}
\psi '\left( {\frac{1}{2} + x} \right) = \frac{1}{x} - \frac{1}{{12{x^3}}} + O\left( {\frac{1}{{{x^4}}}} \right),
\end{equation}
\begin{equation}
\label{psi}
\psi \left( x \right) = \ln x - \frac{1}{{2x}} + O\left( {\frac{1}{{{x^2}}}} \right),
\end{equation}
\begin{equation}
\label{psi1}
\psi '\left( x \right) = \frac{1}{x} + \frac{1}{{2{x^2}}} + O\left( {\frac{1}{{{x^3}}}} \right).
\end{equation}

Finally, we present the Galitski-Larkin integral, which is used for calculations
\begin{equation}
\label{GL_integral}
\int\limits_0^\infty  {\frac{{dx}}{{{{\sinh }^2}x}}} \frac{{{x^2}}}{{{a^2} + {x^2}}} = \frac{a}{\pi }\psi '\left( {\frac{a}{\pi }} \right) - \frac{\pi }{{2a}} - 1,
\end{equation}
where $a$ is an arbitrary value.

\subsection{Transfer from momentum to Landau representation} \label{MtoL}

Using the relation $\mathcal{D}q^{2}\rightarrow\omega_{c}\left(m+\frac{1}{2}\right)$
to substitute the summation over the 2D momentum by that over the
degenerate states of each Landau level:

\begin{equation}
\mathcal{D}\int\frac{d^{2}q}{\left(2\pi\right)^{2}}f(q^{2})=\frac{\mathcal{D}}{4\pi}\int dq^{2}f(q^{2})\rightarrow\frac{\omega_{c}}{4\pi}\sum_{m=0}^{\infty}f(m).
\end{equation}

At zero temperature close to $H_{c2}(0)$, one can perform the calculations
in the LLL approximation:
\begin{equation}
\frac{\omega_{c}(H)}{4\pi}\sum_{m=0}^{\infty}f(m)\approx\frac{\omega_{c}(H_{c2})}{4\pi}f(m=0)=\frac{\Delta}{2\pi}f(m=0).
\end{equation}

\subsection{Eliashberg transformation}
\begin{equation}
\sum_n f(n)=\frac{1}{2i}\oint\coth\left(\pi z\right)f(-iz)dz \label{Eliashberg}
\end{equation}

\section{Aslamazov-Larkin contribution}
\subsection{Basic Expressions}

Let us start from discussion of the AL contribution (diagram 1 at
the Fig. 2 of main text). Corresponding analytic
expression is
\begin{align}
Q_{xx}^{AL}(\omega_{\nu})= -4e^{2}T\sum_{\Omega_{k}}\left(\frac{eH}{\pi}\right)\sum_{\{n,m\}=0}^{\infty}B_{n,m}(\Omega_{k}+\omega_{\nu},\Omega_{k})L_{m}(\Omega_{k})B_{m,n}(\Omega_{k},\Omega_{k}+\omega_{\nu})L_{n}(\Omega_{k}+\omega_{\nu}). \label{sigfield}
\end{align}
where we added one density of states $eH/\pi$ at the Landau level.
Not two because of the orthogonality. The block of three Green functions
with the velocity operator (originating from the current vertex) and
two Cooperons
\begin{equation}
B_{n,m}(\Omega_{k}+\omega_{\nu},\Omega_{k})=T\sum_{\varepsilon_{n}}\mathrm{Tr}\left\{ G\left(\varepsilon_{n}\right)\widehat{v}G\left(\varepsilon_{n}+\omega_{\nu}\right)\widehat{\lambda}(\varepsilon_{n}+\omega_{\nu},\Omega_{k}-\varepsilon_{n})G\left(\Omega_{k}-\varepsilon_{n}\right)\widehat{\lambda}(\Omega_{k}-\varepsilon_{n},\varepsilon_{n})\right\} \label{b}
\end{equation}
was calculated in \cite{SM-GL2001} exactly for the fields with $\omega_{\mathrm{c}}\tau\ll1,$
i.e. namely the case of our interest. At this condition the Landau
quantization concerns the motion of Cooper pairs, while the Green
functions in the block Eq. (\ref{b}) can be taken in $\tau-$approximation.
In result, using the properties of velocity operator in Landau representation,
one finds
\begin{equation}
B_{n,m}(\Omega_{k}+\omega_{\nu},\Omega_{k})=-2\nu_{0}\mathcal{D}\left[\sqrt{eH(n+1)}\delta_{m,n+1}+\sqrt{eHn}\delta_{m,n-1}\right]\Xi_{n,m}(\Omega_{k}+\omega_{\nu},\Omega_{k}),\label{Bsi}
\end{equation}
where
\begin{equation}
\Xi_{n,m}(\Omega_{k}+\omega_{\nu},\Omega_{k})=2\pi T\sum_{\varepsilon_{i}}\frac{\Theta\left(-\left(\varepsilon_{i}+\omega_{\nu}\right)\left(\Omega_{k}-\varepsilon_{i}\right)\right)}{|2\varepsilon_{i}+\omega_{\nu}-\Omega_{k}|+\omega_{\mathrm{c}}(n+1/2)}\frac{\Theta\left(-\varepsilon_{i}\left(\Omega_{k}-\varepsilon_{i}\right)\right)}{|2\varepsilon_{i}-\Omega_{k}|+\omega_{\mathrm{c}}(m+1/2)}.\label{thesi}
\end{equation}
Substitution of Eq. (\ref{Bsi}) to Eq. (\ref{sigfield}) and further
summation over Landau levels\ in Eq. (\ref{sigfield}) results in
cancellation of the terms containg the products $\delta_{m,n+1}\delta_{n,m+1}$
and $\delta_{m,n-1}\delta_{n,m-1}.$ Now we have to study the properties
of $\Xi_{mn}(\Omega_{k},\Omega_{k}+\omega_{\nu})$. Analysis of theta-functions
in Eq. (\ref{thesi}) results in
\begin{align}
\Xi_{mn}(\Omega_{k},\Omega_{k}+\omega_{\nu})= & 2\pi T\left[\Theta\left(\Omega_{k}\right)\sum_{i=k}^{\infty}+\Theta\left(-\Omega_{k}\right)\sum_{i=0}^{\infty}+\Theta\left(-\Omega_{k}-\omega_{\nu}\right)\sum_{i=-\infty}^{k-1}+\Theta\left(\Omega_{k}+\omega_{\nu}\right)\sum_{i=-\infty}^{\nu-1}\right]\cdot\nonumber \\
 & \frac{1}{|2\varepsilon_{i}+\omega_{\nu}-\Omega_{k}|+\omega_{\mathrm{c}}(n+1/2)}\frac{1}{|2\varepsilon_{i}-\Omega_{k}|+\omega_{\mathrm{c}}(m+1/2)}.\label{sigma}
\end{align}
Summation over fermionic frequency in this expression can be performed
in terms of $\psi-$functions:
\begin{align}
\Xi_{mn}(\Omega_{k},\Omega_{k}+\omega_{\nu})= & \frac{1}{2\omega_{\mathrm{c}}\left(n-m\right)}\left[\psi\left(\frac{1}{2}+\frac{\omega_{\nu}+|\Omega_{k}|+\omega_{\mathrm{c}}(m+1/2)}{4\pi T}\right)-\psi\left(\frac{1}{2}+\frac{|\Omega_{k}|+\omega_{\mathrm{c}}(m+1/2)}{4\pi T}\right)\right.\nonumber \\
 & +\left.\psi\left(\frac{1}{2}+\frac{|\Omega_{k+\nu}|+\omega_{\mathrm{c}}(m+1/2)}{4\pi T}\right)-\psi\left(\frac{1}{2}+\frac{\omega_{\nu}+|\Omega_{k+\nu}|+\omega_{\mathrm{c}}(m+1/2)}{4\pi T}\right)\right].\label{gransig-1}
\end{align}
Being interested in the fluctuation conductivity at frequencies $\omega\approx\Delta\tilde h \ll \Delta$,
we omitted in Eq. (\ref{gransig-1}) frequency $\omega_{\nu}$ in
comparison to $\omega_{\mathrm{c}}\left(n-m\right)$ in denominator
since the diagonal term ($m=n$) in process of summation over Landau
levels in Eq. (\ref{sigfield}) disappears as it follows from Eq.
(\ref{Bsi})$.$

One can see that the permutation $\Omega_{k}\ \Leftrightarrow\Omega_{k}+\omega_{\nu}$
simultaneously with $m\Leftrightarrow n$ in Eq. (\ref{gransig-1})
does not change the function $\Xi_{mn}(\Omega_{k},\Omega_{k}+\omega_{\nu}):$
\begin{equation}
\Xi_{mn}(\Omega_{k},\Omega_{k}+\omega_{\nu})\equiv\Xi_{nm}(\Omega_{k}+\omega_{\nu},\Omega_{k}).\label{permut-1}
\end{equation}

Let us return to the general expression for paraconductivity Eq. (\ref{sigfield})
.One can transform the sum over the bosonic frequency $\Omega_{k}$
to the contour integral $I^{AL}$ in the plane of complex frequency
$\Omega_{k}\rightarrow-iz$:
\begin{equation}
Q_{xx}^{AL}(\omega_{\nu})=-\frac{16}{\pi}e^{4}\nu_{0}^{2}\mathcal{D}^{2}H^{2}\sum_{\{n,m\}}^{\infty}C_{mn}I_{nm}^{AL}\left(\omega_{\nu}\right),\label{qan-2}
\end{equation}
\begin{equation}
I_{nm}^{AL}\left(\omega_{\nu}\right)=\frac{1}{4\pi i}{\displaystyle \oint}\coth\left(\frac{z}{2T}\right)dz\Xi_{nm}(-iz+\omega_{\nu},-iz)\Xi_{mn}(-iz,-iz+\omega_{\nu})L_{m}(-iz)L_{n}(-iz+\omega_{\nu}), \label{eliashx}
\end{equation}
where the contour integral encloses all frequencies $\Omega_{k}$
(in the plane of frequency $z$ these are poles of $\coth\left(z/2T\right)$).
The coefficients
\begin{equation}
C_{mn}=\delta_{m,n+1}\delta_{n,m-1}\sqrt{m}\sqrt{n+1}+\delta_{n,m+1}\delta_{m,n-1}\sqrt{n}\sqrt{m+1}\label{c-2}
\end{equation}
govern summation over Landau levels.

\begin{figure}
	\includegraphics[width=0.5\columnwidth]{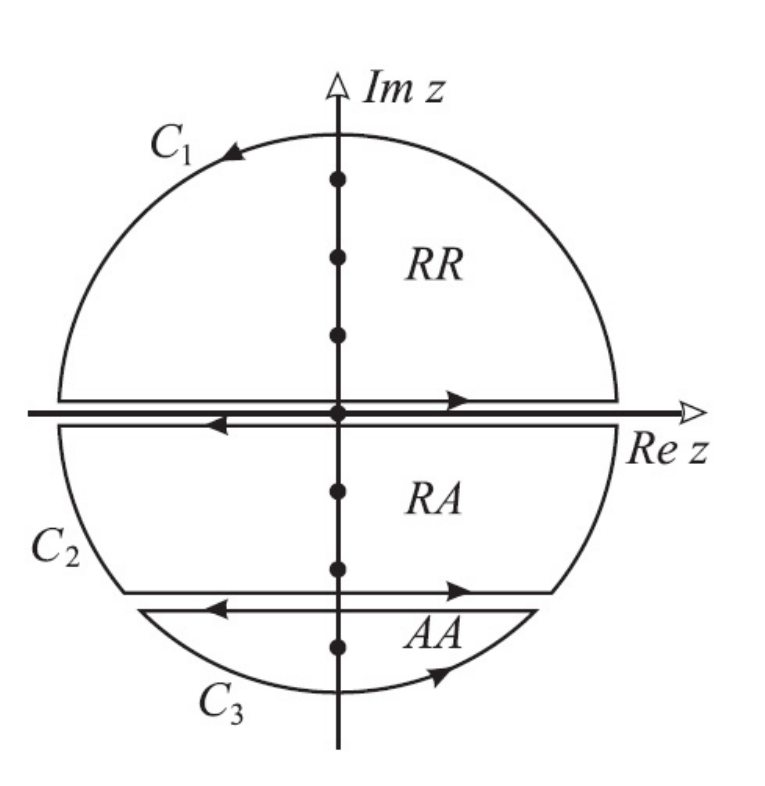}
	\caption {The integration contour in the plane of complex frequencies.}
\label{AL_contour}
\end{figure}

One can notice that in Eq. (\ref{eliashx}) both $\Xi$ functions
have breaks of their analyticity along the lines $\Im z=0$ and $\Im z=-\omega_{\nu}$ (see Fig. \ref{AL_contour}),
the same as the product of propagators. In result one gets three domains
where the integrand function is analytical:\ above the line $Imz=0$,
between the lines $Imz=0$ and $Imz=-\omega_{\nu}$ and below $Imz=-\omega_{\nu}.$
As the analytical continuation of function (\ref{gransig-1}) on the
whole complex plane from Matsubara frequencies three different functions:
$\Xi_{nm}^{RR},\Xi_{nm}^{RA}$ and $\Xi_{nm}^{AA},$ analytical in
corresponding domain, should be introduced. They differ by the combinations
of the signs in the modulus of Eq. (\ref{gransig-1}). Due to observation
Eq. (\ref{permut-1}) one can write the useful identities
\begin{align}
\Xi_{nm}^{RR}(-iz+\omega_{\nu},-iz)= & \Xi_{mn}^{RR}(-iz,-iz+\omega_{\nu}),\\
\Xi_{nm}^{AA}(-iz^{\prime},-iz^{\prime}-\omega_{\nu})= & \Xi_{mn}^{AA}(-iz^{\prime}-\omega_{\nu},-iz^{\prime}),\\
\Xi_{nm}^{RA}(-iz+\omega_{\nu},-iz)= & \Xi_{mn}^{RA}(-iz,-iz+\omega_{\nu}),
\end{align}
and get for the contour integral in Eq. (\ref{eliashx}) :
\begin{align}
& 4\pi iI_{nm}^{AL}\left(\omega_{\nu}\right) =\int_{-\infty}^{\infty}\coth\left(\frac{z}{2T}\right)dz\left\{ \left[\Xi_{nm}^{RR}(-iz+\omega_{\nu},-iz)\right]^{2}L_{m}^{R}(-iz)-\left[\Xi_{nm}^{RA}(-iz+\omega_{\nu},-iz)\right]^{2}L_{m}^{A}(-iz)\right\} L_{n}^{R}(-iz+\omega_{\nu}) \nonumber \\
 & +\int_{-\infty-i\omega_{\nu}}^{\infty-i\omega_{\nu}}\coth\left(\frac{z}{2T}\right)dz\left\{ \left[\Xi_{nm}^{RA}(-iz+\omega_{\nu},-iz)\right]^{2}L_{n}^{R}(-iz+\omega_{\nu})-\left[\Xi_{nm}^{AA}(-iz+\omega_{\nu},-iz)\right]^{2}L_{n}^{A}(-iz+\omega_{\nu})\right\} L_{m}^{A}(-iz).
 \label{I_AL_mn}
\end{align}

The last integral can be reduced to the integration along the real
axis by means of shifting the variable $-iz+\omega_{\nu}\rightarrow-iz^{\prime}.$
In result the expression (\ref{eliashx}) \ for electromagnetic response
operator, still defined on Matsubara frequencies $\omega_{\nu},$
takes form:
\begin{equation}
Q_{xx}^{AL}(\omega_{\nu})=-\frac{16}{\pi}e^{4}\nu_{0}^{2}\mathcal{D}^{2}H^{2}\sum_{n,m}^{\infty}C_{mn}\frac{1}{4\pi i}\int_{-\infty}^{\infty}\coth\left(\frac{z}{2T}\right)\Phi_{mn}\left(z,\omega_{\nu}\right)dz \label{eliash1-2}
\end{equation}
where
\begin{align*}
\Phi_{mn}\left(z,\omega_{\nu}\right) & =\left\{ \left[\Xi_{nm}^{RR}(-iz+\omega_{\nu},-iz)\right]^{2}L_{m}^{R}(-iz)-\left[\Xi_{nm}^{RA}(-iz+\omega_{\nu},-iz)\right]^{2}L_{m}^{A}(-iz)\right\} L_{n}^{R}(-iz+\omega_{\nu})\\
 & +\left\{ \left[\Xi_{mn}^{RA}(-iz-\omega_{\nu},-iz)\right]^{2}L_{n}^{R}(-iz)-\left[\Xi_{nm}^{AA}(-iz,-iz-\omega_{\nu})\right]^{2}L_{n}^{A}(-iz)\right\} L_{m}^{A}(-iz-\omega_{\nu}).
\end{align*}
The rules of writing down the analytical continuations of the function $\Xi_{mn}(\Omega_{k},\Omega_{k}+\omega_{\nu})$ are simple: the sign modulus of the corresponding
frequency in Eq. (\ref{gransig-1}) is chosen as "$+$" in the case
of retarded continuation (superscript R) and it is chosen as "$-$"
in the case of advanced one (superscript A). For instance
\begin{align}
\Xi_{mn}^{RA}(\Omega_{k},\Omega_{k}+\omega_{\nu})= & \frac{1}{2\omega_{\mathrm{c}}\left(n-m\right)}\left[\psi\left(\frac{1}{2}+\frac{\omega_{\nu}-\Omega_{k}+\omega_{\mathrm{c}}(n+1/2)}{4\pi T}\right)-\psi\left(\frac{1}{2}+\frac{-\Omega_{k}+\omega_{\mathrm{c}}(m+1/2)}{4\pi T}\right)\right.\nonumber \\
 & +\left.\psi\left(\frac{1}{2}+\frac{\omega_{\nu}+\Omega_{k}+\omega_{\mathrm{c}}(n+1/2)}{4\pi T}\right)-\psi\left(\frac{1}{2}+\frac{2\omega_{\nu}+\Omega_{k}+\omega_{\mathrm{c}}(m+1/2)}{4\pi T}\right)\right]\label{Sig-1}
\end{align}
and analogously for $\Xi_{nm}^{RR}$ and $\Xi_{nm}^{AA}.$

\subsection{AL contribution close to $H_{c2}(0)$ (LLL approximation).}
Below we will be interested only in the region of quantum fluctuation,
i.e. one of the propagators must be taken at zero Landau level.

\begin{align}
Q_{xx}^{AL}(\omega_{\nu}) & =-\frac{16}{\pi}e^{4}\nu_{0}^{2}\mathcal{D}^{2}H^{2}\left[C_{01}I_{10}^{AL}\left(\omega_{\nu}\right)+C_{10}I_{01}^{AL}\left(\omega_{\nu}\right)\right] \nonumber \\
 & =\frac{ie^{2}\nu_{0}^{2}\Delta^{2}}{\pi^{2}}\int_{-\infty}^{\infty}dz\coth\left(\frac{z}{2T}\right)\left[\Phi_{01}\left(z,\omega_{\nu}\right)+\Phi_{10}\left(z,\omega_{\nu}\right)\right]
 \label{qan-1-1}
\end{align}

Using the symmetry of subscripts and arguments permutation we find

\begin{align}
\Phi_{01}\left(z,\omega_{\nu}\right)+\Phi_{10}\left(z,\omega_{\nu}\right) & =\left[\Xi_{10}^{RR}(-iz+\omega_{\nu},-iz)\right]^{2}L_{0}^{R}(-iz)L_{1}^{R}(-iz+\omega_{\nu})+\left[\Xi_{01}^{RR}(-iz+\omega_{\nu},-iz)\right]^{2}L_{1}^{R}(-iz)L_{0}^{R}(-iz+\omega_{\nu})\nonumber \\
 & -\left[\Xi_{10}^{RA}(-iz+\omega_{\nu},-iz)\right]^{2}L_{0}^{A}(-iz)L_{1}^{R}(-iz+\omega_{\nu})-\left[\Xi_{01}^{RA}(-iz+\omega_{\nu},-iz)\right]^{2}L_{1}^{A}(-iz)L_{0}^{R}(-iz+\omega_{\nu})\nonumber \\
 & +\left[\Xi_{10}^{RA}(-iz,-iz-\omega_{\nu})\right]^{2}L_{1}^{R}(-iz)L_{0}^{A}(-iz-\omega_{\nu})+\left[\Xi_{01}^{RA}(-iz,-iz-\omega_{\nu})\right]^{2}L_{0}^{R}(-iz)L_{1}^{A}(-iz-\omega_{\nu})\nonumber \\
 & -\left[\Xi_{10}^{AA}(-iz,-iz-\omega_{\nu})\right]^{2}L_{1}^{A}(-iz)L_{0}^{A}(-iz-\omega_{\nu})-\left[\Xi_{01}^{AA}(-iz,-iz-\omega_{\nu})\right]^{2}L_{0}^{A}(-iz)L_{1}^{A}(-iz-\omega_{\nu})\label{fgran-1-1-1-2}
\end{align}

Let us recall that $\omega_{c}=2\Delta$ and 
\[
L_{m}^{-1}\left(\Omega_{k}\right)=-\nu_{0}\left[\ln\left(2m+1\right)+\widetilde{h}+\frac{|\Omega_{k}|}{\left(2m+1\right)\Delta}\right],
\]
hence in the assumption that $z,\omega\ll\Delta$ one can accept
\[
L_{1}^{R}(-iz)=L_{1}^{R}(-iz+\omega_{\nu})=L_{1}^{A}(-iz)=L_{1}^{A}(-iz-\omega_{\nu})=-\frac{1}{\nu_{0}\ln3}.
\]
Hence
\begin{align}
-\nu_{0}\ln3\left[\Phi_{01}\left(z,\omega_{\nu}\right)+\Phi_{10}\left(z,\omega_{\nu}\right)\right] & =\left\{ \left[\Xi_{10}^{RR}(-iz+\omega_{\nu},-iz)\right]^{2}+\left[\Xi_{01}^{RA}(-iz,-iz-\omega_{\nu})\right]^{2}\right\} L_{0}^{R}(-iz)\nonumber \\
 & -\left\{ \left[\Xi_{01}^{AA}(-iz,-iz-\omega_{\nu})\right]^{2}+\left[\Xi_{10}^{RA}(-iz+\omega_{\nu},-iz)\right]^{2}\right\} L_{0}^{A}(-iz)\nonumber \\
 & +\left\{ \left[\Xi_{10}^{RA}(-iz,-iz-\omega_{\nu})\right]^{2}-\left[\Xi_{10}^{AA}(-iz,-iz-\omega_{\nu})\right]^{2}\right\} L_{0}^{A}(-iz-\omega_{\nu})\nonumber \\
 & +\left\{ \left[\Xi_{01}^{RR}(-iz+\omega_{\nu},-iz)\right]^{2}-\left[\Xi_{01}^{RA}(-iz+\omega_{\nu},-iz)\right]^{2}\right\} L_{0}^{R}(-iz+\omega_{\nu})\label{fgran-1-1-1-1-1}
\end{align}

\subsection{Frequency independent AL contribution to conductivity.}

The elements in first two lines of Eq. (\ref{fgran-1-1-1-1-1}) have the asymptotic behaviors ($\psi(1/2+x)\approx\ln x+\frac{1}{24x^{2}}$) near
zero temperatures:
\begin{align*}
\Xi_{10}^{RR}(-iz+\omega_{\nu},-iz) & \approx\frac{1}{4\Delta}\ln\left(\frac{\left(\omega_{\nu}-iz+3\Delta\right)^{2}}{\left(-iz+\Delta\right)\left(-iz+2\omega_{\nu}+\Delta\right)}\right),\\
\Xi_{01}^{RA}(-iz,-iz-\omega_{\nu}) & \approx\frac{1}{4\Delta}\ln\left(\frac{\left(3\Delta\right)^{2}-\left(\omega_{\nu}+iz\right)^{2}}{\Delta^{2}+z^{2}}\right),\\
\Xi_{01}^{AA}(-iz,-iz-\omega_{\nu}) & \approx\frac{1}{4\Delta}\ln\left(\frac{\left(\omega_{\nu}+iz+3\Delta\right)^{2}}{\left(iz+\Delta\right)\left(iz+2\omega_{\nu}+\Delta\right)}\right),\\
\Xi_{10}^{RA}(-iz+\omega_{\nu},-iz) & \approx\frac{1}{4\Delta}\ln\left(\frac{\left(3\Delta\right)^{2}-\left(\omega_{\nu}-iz\right)^{2}}{\Delta^{2}+z^{2}}\right).
\end{align*}
Then performing the analytical continuation $\omega_{\nu}\rightarrow-i\omega$ and ignoring the terms of the order of $\omega\sim\Delta$, one has
\begin{eqnarray}
\left[\Xi_{10}^{RR}(-iz-i\omega,-iz)\right]^{2}+\left[\Xi_{01}^{RA}(-iz,-iz+i\omega)\right]^{2} & \approx\frac{\ln3}{2\Delta^{2}}\left[\ln3+\left(\frac{2i}{3}-\frac{z}{\Delta}\right)\frac{\omega}{\Delta}\right]
\label{sig1}
\end{eqnarray}
\begin{eqnarray}
\left[\Xi_{01}^{AA}(-iz,-iz+i\omega)\right]^{2}+\left[\Xi_{10}^{RA}(-iz-i\omega,-iz)\right]^{2} & \approx\frac{\ln3}{2\Delta^{2}}\left[\ln3+\left(\frac{2i}{3}+\frac{z}{\Delta}\right)\frac{\omega}{\Delta}\right].
\label{sig2}
\end{eqnarray}

Substituting Eqs. (\ref{sig1})-(\ref{sig2}) into Eq. (\ref{fgran-1-1-1-1-1}), one obtains the contribution to the paraconductivity which does not depend on frequency in the range of $\omega \ll \Delta$:
\begin{align}
Q_{1.1}^{AL}(\omega) & =-\frac{ie^{2}\nu_{0}\Delta^{2}}{\pi^{2}\ln3}\int_{-\infty}^{\infty}dz\coth\left(\frac{z}{2T}\right)\left(\left\{ \left[\Xi_{10}^{RR}(-iz-i\omega,-iz)\right]^{2}+\left[\Xi_{01}^{RA}(-iz,-iz+i\omega)\right]^{2}\right\} L_{0}^{R}(-iz)\right. \nonumber \\
 &=i\frac{2e^{2}\omega}{3\pi^{2}}\left(2-3\widetilde{h}\right)\left(\frac{T}{\Delta}\right)^{2}\int_{-\infty}^{\infty}dz\frac{z\coth z}{\widetilde{h}^{2}+\left(\frac{2T}{\Delta}\right)^{2}z^{2}},
\label{Q11}
\end{align}
\begin{align}
Q_{1.2}^{AL}(\omega) & =\frac{ie^{2}\nu_{0}\Delta^{2}}{\pi^{2}\ln3}\int_{-\infty}^{\infty}dz\coth\left(\frac{z}{2T}\right)\left\{ \left[\Xi_{01}^{AA}(-iz,-iz-\omega_{\nu})\right]^{2}+\left[\Xi_{10}^{RA}(-iz+\omega_{\nu},-iz)\right]^{2}\right\} L_{0}^{A}(-iz) \nonumber\\
 & =-i\frac{2e^{2}\omega}{3\pi^{2}}\left(2+3\widetilde{h}\right)\left(\frac{T}{\Delta}\right)^{2}\int_{-\infty}^{\infty}dz\frac{z\coth z}{\widetilde{h}^{2}+\left(\frac{2T}{\Delta}\right)^{2}z^{2}}.
\label{Q12}
\end{align}
Performing the integrations and using relation Eq. (\ref{Delta_Tc}), one gets
\begin{equation}
Q_{1}^{AL}(\omega)-Q_{1}^{AL}(0)=Q_{1.1}^{AL}(\omega)+Q_{1.2}^{AL}(\omega)=-i\omega\frac{2e^{2}}{\pi^{2}}\widetilde{h}\left[\ln\frac{1}{\widetilde{h}}+\frac{1}{3}\left(\frac{\gamma_{E}t}{\widetilde{h}}\right)^{2}\right],
\label{difQ}
\end{equation}
and finally
\begin{equation}
\sigma_{1}^{AL}=\frac{2e^{2}}{\pi^{2}}\widetilde{h}\left[\ln\frac{1}{\widetilde{h}}+\frac{1}{3}\left(\frac{\gamma_{E}t}{\widetilde{h}}\right)^{2}\right].
\label{sigma1_AL}
\end{equation}

\subsection{Frequency dependent AL contribution to the conductivity}
Let us proceed to the analysis of the contribution to the paraconductivity emerging from the third and fourth lines of Eq. (\ref{fgran-1-1-1-1-1}):
\begin{align}
Q_{2}^{AL}(\omega_{\nu}) = &-\frac{ie^{2}\nu_{0}\Delta^{2}}{\pi^{2}\ln3}\int_{-\infty}^{\infty}dz\coth\left(\frac{z}{2T}\right)\left(\left\{ \left[\Xi_{10}^{RA}(-iz,-iz-\omega_{\nu})\right]^{2}-\left[\Xi_{10}^{AA}(-iz,-iz-\omega_{\nu})\right]^{2}\right\} L_{0}^{A}(-iz-\omega_{\nu})\right. \nonumber \\& +\left.\left\{ \left[\Xi_{01}^{RR}(-iz+\omega_{\nu},-iz)\right]^{2}-\left[\Xi_{01}^{RA}(-iz+\omega_{\nu},-iz)\right]^{2}\right\} L_{0}^{R}(-iz+\omega_{\nu})\right) \nonumber \\
 = & Q_{2,1}^{AL}(\omega_{\nu})+Q_{2,2}^{AL}(\omega_{\nu}),
 \label{Q2_AL_gen}
\end{align}
where
\begin{align}
\Xi_{10}^{RA}(-iz,-iz-\omega_{\nu}) & \approx-\frac{1}{4\Delta}\ln\left(\frac{\Delta^{2}-\left(iz+\omega_\nu\right)^{2}}{\left(3\Delta\right)^{2}+z^{2}}\right),\\
\Xi_{10}^{AA}(-iz,-iz-\omega_{\nu}) & \approx-\frac{1}{4\Delta}\ln\left(\frac{\left(iz+\omega_\nu+\Delta\right)^{2}}{\left(iz+3\Delta\right)\left(2\omega_\nu+iz+3\Delta\right)}\right),\\
\Xi_{01}^{RR}(-iz+\omega_{\nu},-iz) & \approx\frac{1}{4\Delta}\ln\left(\frac{\left(2\omega_\nu-iz+3\Delta\right)\left(-iz+3\Delta\right)}{\left(\omega_\nu-iz+\Delta\right)^{2}}\right),\\
\Xi_{01}^{RA}(-iz+\omega_{\nu},-iz) & \approx\frac{1}{4\Delta}\ln\left(\frac{\left(3\Delta\right)^{2}+z^{2}}{\Delta^{2}-\left(iz-\omega_\nu\right)^{2}}\right).
\label{Xi_10_01_AA_RR}
\end{align}
After the analytical continuations, we expand the logarithms in powers of $(\omega, z)/\Delta$:
\begin{align}
\left[\Xi_{10}^{RA}(-iz,-iz+i\omega)\right]^{2}-\left[\Xi_{10}^{AA}(-iz,-iz+i\omega)\right]^{2} & =\frac{i\ln3}{3\Delta^{3}}\left(z-\omega\right)\left(1-\frac{i\omega}{6\Delta}\right)\left[1-\frac{i\left(z-\omega\right)}{3\ln3\Delta}\right],
\label{Xi_diff_RR_AA1}
\end{align}
\begin{align}
\left[\Xi_{01}^{RR}(-iz-i\omega,-iz)\right]^{2}-\left[\Xi_{01}^{RA}(-iz-i\omega,-iz)\right]^{2} & =\frac{i\ln3\left(z+\omega\right)}{3\Delta^{3}}\left(1-\frac{i\omega}{6\Delta}\right)\left[1+\frac{i\left(z+\omega\right)}{3\ln3\Delta}\right].
\label{Xi_diff_RR_AA2}
\end{align}
Substituting Eqs. (\ref{Xi_diff_RR_AA1}) and (\ref{Xi_diff_RR_AA2}) into Eq. (\ref{qan-1-1}), one can easily find that
\begin{align}
Q_{2.1}^{AL}(\omega)=Q_{2.2}^{AL}(\omega) & =-\frac{e^{2}}{3\pi^{2}\Delta}\left(1-\frac{i\omega}{6\Delta}\right)\int_{-\infty}^{\infty}dz\coth\left(\frac{z+\omega}{2T}\right)\left(1-\frac{iz}{3\ln3\Delta}\right)\frac{z\left(\widetilde{h}-\frac{iz}{\Delta}\right)}{\left(\widetilde{h}\right)^{2}+\left(\frac{z}{\Delta}\right)^{2}}.
\label{Q21_22_AL}
\end{align}

Recalling that $\tilde{h}\ll1$ one can omit the term $\frac{iz}{3\ln3\Delta}$ in parentheses in Eq. (\ref{Q21_22_AL})
\begin{align}
Q_{2}^{AL(R)}(-i\omega) & =-\frac{2e^{2}}{3\pi^{2}}\left(1-\frac{i\omega}{6\Delta}\right)\frac{\left(2T\right)^{2}}{\Delta}\int_{-\infty}^{\infty}dy\left[\coth\left(y-\frac{\omega}{2T}\right)\right]\frac{y\left(\tilde{h}+i\left(\frac{2T}{\Delta}\right)y\right)}{\left(\tilde{h}\right)^{2}+\left(\frac{2T}{\Delta}\right)^{2}y^{2}}= \nonumber \\
= & -\frac{2e^{2}}{3\pi^{2}}\left(1-\frac{i\omega}{6\Delta}\right)\frac{\left(2T\right)^{2}}{\Delta}J,
\label{eliash1-1-1-2-2-1-1-1-1-1-2-1}
\end{align}
where $y=z/2T$ is the new variable and $J$ has the form
\begin{align}
J= \int_{-\infty}^{\infty}dy\left[\coth\left(y-\frac{\omega}{2T}\right)\right]\frac{y\left(\tilde{h}+i\left(\frac{2T}{\Delta}\right)y\right)}{\left(\tilde{h}\right)^{2}+\left(\frac{2T}{\Delta}\right)^{2}y^{2}}.
\label{eliash1-1-1-2-2-1-1-1-1-1-2-1}
\end{align}
The integral given by Eq. (\ref{eliash1-1-1-2-2-1-1-1-1-1-2-1}) can be evaluated by parts. Let us start from the calculus of the antiderivative of the the fraction in the intergand:
\begin{eqnarray}
\int dy\frac{\left(y\tilde{h}+i\left(\frac{2T}{\Delta}\right)y^{2}\right)}{\left(\tilde{h}\right)^{2}+\left(\frac{2T}{\Delta}\right)^{2}y^{2}}=\tilde{h}\left(\frac{\Delta}{2T}\right)^{2}\ln\left(\sqrt{\left(\tilde{h}\right)^{2}+\left(\frac{2T}{\Delta}\right)^{2}y^{2}}\right)+i\left(\frac{\Delta}{2T}\right)^{2}\int dy\left(1-\frac{\tilde{h}^{2}}{\tilde{h}^{2}+y^{2}}\right) \nonumber \\
=\tilde{h}\left(\frac{\Delta}{2T}\right)^{2}\ln\left(\sqrt{\left(\tilde{h}\right)^{2}+\left(\frac{2T}{\Delta}\right)^{2}y^{2}}\right)+i\left(\frac{\Delta}{2T}\right)^{2}\left(y-\tilde{h}\arctan\frac{y}{\tilde{h}}\right).
\label{integrand_trick}
\end{eqnarray}
This trick allows to transform Eq. (\ref{eliash1-1-1-2-2-1-1-1-1-1-2-1}) to the integral
\begin{eqnarray}
J & =\left(\frac{\Delta}{2T}\right)^{2}\left[\tilde{h}\ln\left(\sqrt{\left(\tilde{h}\right)^{2}+\left(\frac{2T}{\Delta}\right)^{2}y^{2}}\right)+i\left(y\frac{2T}{\Delta}-\tilde{h}\arctan\frac{2Ty}{\Delta\tilde{h}}\right)\right]\coth\left(y-\frac{\omega}{2T}\right)|_{-A}^{A} \nonumber \\
+ & \left(\frac{\Delta}{2T}\right)^{2}\int_{A}^{A}\left(\tilde{h}\ln\left(\sqrt{\left(\tilde{h}\right)^{2}+\left(\frac{2T}{\Delta}\right)^{2}y^{2}}\right)+i\left(y\frac{2T}{\Delta}-\tilde{h}\arctan\frac{2Ty}{\Delta\tilde{h}}\right)\right)\frac{dy}{\sinh^{2}\left(y-\frac{\omega}{2T}\right)}.
\end{eqnarray}

Again, considering that $A=\Delta/2T \gg 1$, $t \ll \tilde{h}$, and  $\tilde{h}\ll1$ and expending in series the square root one obtains:
\begin{align}
\frac{\left(2T\right)^{2}}{\Delta}J & =i\Delta\left[\coth\left(\Delta-\omega\right)/2T-\coth\left(\Delta+\omega\right)/2T\right] \nonumber \\
+ & \Delta\int_{-A}^{A}\left(\tilde{h}\ln\tilde{h}+2\tilde{h}\left(\frac{T}{\Delta\tilde{h}}\right)^{2}y^{2}+i\left(\frac{2Ty}{\Delta}-\tilde{h}\arctan\frac{2Ty}{\Delta\tilde{h}}\right)\right)\frac{dy}{\sinh^{2}\left(y-\frac{\omega}{2T}\right)}.
\label{J_new}
\end{align}
\subsubsection{Limit $\omega\ll T$}
When $\omega\ll T$ one can expand in Eq. (\ref{J_new}) also the arctan function up to the third order
\begin{align}
\frac{\left(2T\right)^{2}}{\Delta}J & =\Delta\int_{-A}^{A}\left(\tilde{h}\ln\tilde{h}+2\tilde{h}\left(\frac{T}{\Delta\tilde{h}}\right)^{2}y^{2}+\frac{i}{3\tilde{h}^{2}}\left(\frac{2Ty}{\Delta}\right)^{3}\right)\left[\frac{dy}{\sinh^{2}y}+\frac{\omega}{T}\frac{\coth y}{\sinh^{2}y}dy\right]\\
= & \Delta\left(2\tilde{h}\ln\tilde{h}+\frac{2\pi^{2}\tilde{h}}{3}\left(\frac{T}{\Delta\tilde{h}}\right)^{2}\right)+\frac{2i\omega}{3\tilde{h}^{2}}\left(\frac{2T}{\Delta}\right)^{2}\int_{-\infty}^{\infty}\frac{y^{3}\coth y}{\sinh^{2}y}dy
\label{J_new2}
\end{align}
From Eq. (\ref{J_new2}) finally one finds:
\begin{align}
\frac{\left(2T\right)^{2}}{\Delta}J & =\Delta\left(2\tilde{h}\ln\tilde{h}+\frac{2\pi^{2}\tilde{h}}{3}\left(\frac{T}{\Delta\tilde{h}}\right)^{2}\right)+\frac{2i\omega}{3\tilde{h}^{2}}\left(\frac{2T}{\Delta}\right)^{2}\pi^2/\textcolor{red}{2}.
\label{J_final}
\end{align}
Based on the result of integration Eq. (\ref{J_final}) one can express 
\begin{align}
Q_{2}^{AL(R)}(-i\omega) & =-\frac{2e^{2}}{3\pi^{2}}\left(1-\frac{i\omega}{6\Delta}\right)\frac{\left(2T\right)^{2}}{\Delta}J=-\frac{2e^{2}}{3\pi^{2}}\left(1-\frac{i\omega}{6\Delta}\right)\left[\Delta\left(2\tilde{h}\ln\tilde{h}+\frac{2\pi^{2}\tilde{h}}{3}\left(\frac{T}{\Delta\tilde{h}}\right)^{2}\right)+\frac{i\omega}{3\tilde{h}^{2}}\left(\frac{2T}{\Delta}\right)^{2}\pi^2
\right],
\label{Q2_AL_R_new}
\end{align}
which finally gives for the $Q_2(-i\omega)-Q_2(0)$ operator 
\begin{align}
Q_{2}^{AL(R)}(-i\omega)-Q_{2}^{AL(R)}(0)=-\frac{2e^{2}}{3\pi^{2}}\left(-\frac{i\omega}{6\Delta}\right)\left[\Delta\left(2\tilde{h}\ln\tilde{h}+\frac{2\pi^{2}\tilde{h}}{3}\left(\frac{T}{\Delta\tilde{h}}\right)^{2}\right)+\frac{i\omega}{3\tilde{h}^{2}}\left(\frac{2T}{\Delta}\right)^{2}\pi^{2}\right]-\frac{2e^{2}}{3\pi^{2}}\left[\frac{i\omega}{3\tilde{h}^{2}}\left(\frac{2T}{\Delta}\right)^{2}\pi^{2}\right]
\label{Q2_AL_R_new}
\end{align}
\begin{align}
Q_{2}^{AL(R)}(-i\omega)-Q_{2}^{AL(R)}(0)=+\frac{e^{2}}{9\pi^{2}}i\omega\left[\left(2\tilde{h}\ln\tilde{h}+\frac{2\pi^{2}\tilde{h}}{3}\left(\frac{T}{\Delta\tilde{h}}\right)^{2}\right)+\frac{i\omega}{3\tilde{h}^{2}\Delta}\left(\frac{2T}{\Delta}\right)^{2}\pi^{2}-\frac{2\pi^{2}}{\tilde{h}^{2}}\left(\frac{2T}{\Delta}\right)^{2}\right],
\label{Q2_AL_R_new2}
\end{align}
\begin{align}
Q_{2}^{AL(R)}(-i\omega)-Q_{2}^{AL(R)}(0)=+\frac{e^{2}}{9\pi^{2}}i\omega\left[2\tilde{h}\ln\tilde{h}-8\left(\frac{\gamma_{E}t}{\tilde{h}}\right)^{2}+\frac{2\tilde{h}}{3}\left(\frac{\gamma_{E}t}{\tilde{h}}\right)^{2}+\frac{4i\omega}{3\Delta}\left(\frac{\gamma_{E}t}{\tilde{h}}\right)^{2}\right],
\end{align}
corresponding contribution to the conductivity is represented by
\begin{align}
\sigma_{2}^{AL}(\omega)=\frac{e^{2}}{9\pi^{2}}\left[8\left(\frac{\gamma_{E}t}{\tilde{h}}\right)^{2}\left(1-\frac{\tilde{h}}{12}\right)-2\tilde{h}\ln\tilde{h}-\frac{4i\omega}{3\Delta}\left(\frac{\gamma_{E}t}{\tilde{h}}\right)^{2}\right]
\end{align}

The integration gives (Here $\omega$ is arbitrary with respect to T)
\begin{align*}
Q_{2}^{AL}(\omega) & =\frac{4e^{2}}{3\pi^{2}}\left(1-\frac{i\omega}{6\Delta}\right)\left[i\omega+\Delta\widetilde{h}\ln\left(\widetilde{h}-i\frac{\omega}{\Delta}\right)-\frac{\pi^{2}}{3}\frac{T^{2}}{\Delta\widetilde{h}}\left(1+\frac{2i\omega}{\Delta\widetilde{h}}\right)\right],
\end{align*}
Extracting $Q_{2}^{AL}(0)$ one finds
\begin{align*}
Q_{2}^{AL}(\omega)-Q_{2}^{AL}(0)=\frac{4e^{2}}{3\pi^{2}}\left[i\omega+\Delta\widetilde{h}\ln\left(1-i\frac{\omega}{\Delta\widetilde{h}}\right)-\frac{2\pi^{2}i\omega}{3}\left(\frac{T}{\Delta\widetilde{h}}\right)^{2}\right] \\-\frac{2e^{2}}{9\pi^{2}}\frac{i\omega}{\Delta}\left[i\omega+\Delta\widetilde{h}\ln\left(\widetilde{h}-i\frac{\omega}{\Delta}\right)-\frac{\pi^{2}}{3}\frac{T^{2}}{\Delta\widetilde{h}}\left(1+\frac{2i\omega}{\Delta\widetilde{h}}\right)\right],
\end{align*}
what gives the second, frequency dependent,  contribution to paraconductivity
\begin{align*}
\sigma_{2}^{\mathrm{AL}}(\omega)=-\frac{2e^{2}}{9\pi^{2}}\widetilde{h}\ln\frac{1}{\widetilde{h}}+\frac{8e^{2}}{9\pi^{2}}\left(\frac{\gamma_{E}t}{\widetilde{h}}\right)^{2}-\frac{4e^{2}}{3\pi^{2}}\left[1+\frac{\Delta\widetilde{h}}{i\omega}\ln\left(1-\frac{i\omega}{\Delta\widetilde{h}}\right)\right].
\end{align*}
One can see that it is similar to $\sigma^{AL}_1$, but the $\omega$ dependent term appears in it.
Summing up both of them, we arrive to the final result:
\[
\sigma_{xx}^{\mathrm{AL}}(\omega\ll\Delta)=\frac{16e^{2}}{9\pi^{2}}\widetilde{h}\ln\frac{1}{\widetilde{h}}+\frac{8e^{2}}{9\pi^{2}}\left(\frac{\gamma_{E}t}{\widetilde{h}}\right)^{2}-\frac{4e^{2}}{3\pi^{2}}\left[1+\frac{\Delta\widetilde{h}}{i\omega}\ln\left(1-\frac{i\omega}{\Delta\widetilde{h}}\right)\right].
\]

\section{Maki-Thompson Contribution}

\subsection{Basic Expressions}
We start with the usual expression for the Maki-Thompson contribution
written in momentum representation and then, performing integration
over the electronic momentum, we will quantize the motion of Cooper pairs
in magnetic field. The diagram 2 from the Fig. 2 in the main text can be written as
\begin{equation}
Q_{\alpha\beta}^{\mathrm{MT}}(\omega_{\nu})=2e^{2}T\sum_{\Omega_{k}}\int{\frac{{d^{2}}\mathbf{q}}{{(2\pi)^{2}}}}L(\mathbf{q},\Omega_{k})I_{\alpha\beta}^{\mathrm{MT}}(\mathbf{q},\Omega_{k},\omega_{\nu}), \label{d21}
\end{equation}
where
\begin{equation}
I_{\alpha\beta}^{\mathrm{MT}}(\mathbf{q},\Omega_{k},\omega_{\nu})=T\sum_{\varepsilon_{n}}\lambda(\mathbf{q},\varepsilon_{n+\nu},\Omega_{k-n-\nu})\lambda(\mathbf{q},\varepsilon_{n},\Omega_{k-n})J_{\alpha\beta}(\mathbf{q},\varepsilon_{n},\Omega_{k},\omega_{\nu})\label{d22}
\end{equation}
and
\[
J_{\alpha\beta}(\mathbf{q},\varepsilon_{n},\Omega_{k},\omega_{\nu})=\int{\frac{{d^{3}}\mathbf{p}}{{(2\pi)^{3}}}}v_{\alpha}(\mathbf{p})v_{\beta}(\mathbf{q-p})G(\mathbf{p},\varepsilon_{n+\nu})G(\mathbf{p},\varepsilon_{n})G(\mathbf{q-p},\Omega_{k-n-\nu})G(\mathbf{q-p},\Omega_{k-n}).
\]

The main $q$-dependence in (\ref{d21}) arises from the propagator
and vertices $\lambda$. That is why we can assume $q=0$ in the Green
functions and to calculate the electron momentum integral passing,
as usual, to $\xi(\mathbf{p})$ integration:
\begin{equation}
J_{xx}(0,\varepsilon_{n},\Omega_{k},\omega_{\nu})=-\mathcal{D}\tau^{-1}\nu_0\int_{-\infty}^{\infty}\frac{d\xi}{\xi-i\widetilde{\varepsilon}_{n}}\frac{1}{\xi-i\widetilde{\varepsilon}_{n+\nu}}\frac{1}{\xi-i\widetilde{\Omega}_{k-n}}\frac{1}{\xi-i\widetilde{\Omega}_{k-n-\nu}}.\label{integ}
\end{equation}
This integral (\ref{integ}) can be calculated applying Cauchy
theorem. Closing the contour in upper or lower half-plane by the large
semicircle and noticing, that, due to fast decrease of the integrand
function in Eq. (\ref{integ}), the integral over the semicircle turns
zero, one can express $J_{xx}$ in terms of the sum of corresponding
residues. There are 7 different combinations of the pole positions\ with
respect to the real axis in the complex plane of $\xi$, leading to
non-zero results: two realization corresponding to $\Theta\left(-\varepsilon_{n}\varepsilon_{n+\nu}\right)\Theta\left(\Omega_{k-n}\Omega_{k-n-\nu}\right)\neq0,$
one realization corresponding to $\ \Theta\left(-\varepsilon_{n}\varepsilon_{n+\nu}\right)\Theta\left(-\Omega_{k-n}\Omega_{k-n-\nu}\right)\neq0,$
two realization corresponding to $\Theta\left(\varepsilon_{n}\varepsilon_{n+\nu}\right)\Theta\left(\Omega_{k-n}\Omega_{k-n-\nu}\right)\neq0,$
and realization corresponding to $\Theta\left(\varepsilon_{n}\varepsilon_{n+\nu}\right)\Theta\left(-\Omega_{k-n}\Omega_{k-n-\nu}\right)\neq0.$
Calculating the residues for each situation and assuming that $\widetilde{\varepsilon}_{n}=\left(2\tau\right)^{-1}\mathrm{sgn}(\varepsilon_{n})$
(let us recall that we consider the dirty limit $T\ll\tau^{-1})$
one finds:
\begin{align}
J_{xx}(0,\varepsilon_{n},\Omega_{k},\omega_{\nu}) & =2\pi\mathcal{D}\nu_0\tau^{2}\left\{ \left[\Theta\left(-\varepsilon_{n}\varepsilon_{n+\nu}\right)\Theta\left(\Omega_{k-n}\Omega_{k-n-\nu}\right)+\Theta\left(\varepsilon_{n}\varepsilon_{n+\nu}\right)\Theta\left(-\Omega_{k-n}\Omega_{k-n-\nu}\right)\right]\right.\label{Jgen}\\
 & \left.-2\left[\Theta\left(-\varepsilon_{n}\varepsilon_{n+\nu}\right)\Theta\left(-\Omega_{k-n}\Omega_{k-n-\nu}\right)+\Theta\left(\varepsilon_{n}\varepsilon_{n+\nu}\right)\Theta\left(\Omega_{k-n}\Omega_{k-n-\nu}\right)\right]\right\} .\nonumber
\end{align}
Now one should substitute this expression to Eq. (\ref{d22}) and
perform summation over the fermionic frequency. This is a cumbersome
exercise, which, nevertheless, can be followed out analytically.
Let us mention some tricks helping to perform the summations. One
can see, that the simultaneous permutations $n\rightarrow-n$ and
$k\rightarrow-k$ allows to simplify the sums:
\begin{equation}
\begin{split}
I_{xx}^{\mathrm{MT}} & =I_{xx}^{\mathrm{MT(an)}}+I_{xx}^{\mathrm{MT(reg1)}}\\
= & -2\pi\nu_0\mathcal{D}T\left\{ \sum_{n=-\nu}^{-1}\frac{2\Theta\left(-\Omega_{k-n}\Omega_{k-n-\nu}\right)-\Theta\left(\Omega_{k-n}\Omega_{k-n-\nu}\right)}{\left(|\varepsilon_{n+\nu}-\Omega_{k-n-\nu}|+\mathcal{D}q^{2}\right)\left(|\varepsilon_{n}-\Omega_{k-n}|+\mathcal{D}q^{2}\right)}\right.\\
 & +\left.2\sum_{n=0}^{\infty}\frac{2\Theta\left(\Omega_{k-n}\Omega_{k-n-\nu}\right)-\Theta\left(-\Omega_{k-n}\Omega_{k-n-\nu}\right)}{\left(|\varepsilon_{n+\nu}-\Omega_{k-n-\nu}|+\mathcal{D}q^{2}\right)\left(|\varepsilon_{n}-\Omega_{k-n}|+\mathcal{D}q^{2}\right)}\right\}.  \label{Isum}
\end{split}
\end{equation}

After rewriting the absolute values for the Cooperons, the sums can be expressed in terms of $\psi$ functions. One can write the final expression for the first sum as 
\begin{equation}
I_{xx}^{\mathrm{MT(an)}}=-\frac{\mathcal{D}\nu_0\Theta\left(\omega_{\nu-1}-|\Omega_k|\right)}{\omega_{\nu}+\mathcal{D}q^{2}}\left[\psi\left(\frac{1}{2}+\frac{2\omega_{\nu}-|\Omega_{k}|+\mathcal{D}q^{2}}{4\pi T}\right)-\psi\left(\frac{1}{2}+\frac{|\Omega_{k}|+\mathcal{D}q^{2}}{4\pi T}\right)\right]. \label{Ian}
\end{equation}

For the remaining second sum in Eq. (\ref{Isum}), One
can see that in the first term both moduli are positive. In the second
term we can make a trick changing $k\rightarrow-k,$ with the further
change of the order of summation over bosonic frequency. \ The sum
with $\Theta\left(\Omega_{k-n}\Omega_{k-n-\nu}\right)$ can be calculated
in the spirit of Eq. (\ref{Ian}). Regarding the last sum, containing
$\Theta\left(-\Omega_{k-n}\Omega_{k-n-\nu}\right),$ one can find
that it is exactly equal to zero for any $\Omega_{k}.$ Finally
\begin{equation}
I_{xx}^{\mathrm{MT(reg1)}}=-\frac{\mathcal{D}\nu_0}{\omega_{\nu}}\left[\psi\left(\frac{1}{2}+\frac{2\omega_{\nu}+|\Omega_{k}|+\mathcal{D}q^{2}}{4\pi T}\right)-\psi\left(\frac{1}{2}+\frac{|\Omega_{k}|+\mathcal{D}q^{2}}{4\pi T}\right)\right]. \label{Ireg}
\end{equation}

Substituting the explicit Eqs. (\ref{Ian}) and (\ref{Ireg}) in Eq. (\ref{d21}), the analytic expression for the MT contribution to the electromagnetic response tensor can be written as
\[
Q_{xx}^{\mathrm{MT}}(\omega_\nu)=Q_{xx}^{\mathrm{MT(an)}}(\omega_\nu)+Q_{xx}^{\mathrm{MT(reg1)}}(\omega_\nu),
\]
where
\begin{equation}
Q_{xx}^{\mathrm{MT(an)}}(\omega_\nu)=-2e^{2}T\mathcal{D}\nu_{0}\int\frac{d^{2}\mathbf{q}}{(2\pi)^{2}}\frac{1}{\omega_{\nu}+\mathcal{D}q^{2}
}\sum_{|k|=0}^{\nu-1}L\left(\mathbf{q},\Omega_{k}\right)\left[\psi\left(\frac{1}{2}+\frac{2\omega_{\nu}-|\Omega_{k}|+\mathcal{D}q^{2}}{4\pi T }\right)-\psi\left(\frac{1}{2}+\frac{|\Omega_{k}|+\mathcal{D}q^{2}}{4\pi T}\right)\right]
 \label{d21-2}
\end{equation}
and
\begin{equation}
Q_{xx}^{\mathrm{MT(reg1)}}(\omega_{\nu})=-2e^{2}T\frac{\mathcal{D}\nu_0}{\omega_{\nu}}\sum_{\Omega_{k}}\int{\frac{{d^{2}}\mathbf{q}}{{(2\pi)^{2}}}}L(\mathbf{q},\Omega_{k})\left[\psi\left(\frac{1}{2}+\frac{2\omega_{\nu}+|\Omega_{k}|+\mathcal{D}q^{2}}{4\pi T}\right)-\psi\left(\frac{1}{2}+\frac{|\Omega_{k}|+\mathcal{D}q^{2}}{4\pi T}\right)\right].\label{d21-1}
\end{equation}

\subsection{Anomalous MT contribution close to $H_{c2}(0)$ (LLL approximation)}

\begin{figure}
	\includegraphics[width=0.5\columnwidth]{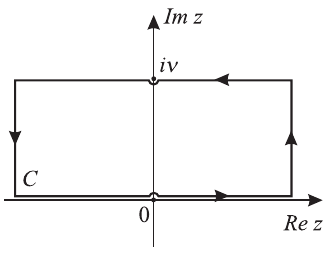}
	\caption {Contour of integration for the MT contribution.}
\label{phasediagram}
\end{figure}

Let us start from the anomalous part (\ref{d21-2}) and transform the momentum integration to the summations over Landau levels (lowest Landau level here),
\[
Q_{xx}^{\mathrm{MT(an)}}(\omega_\nu)=-\frac{e^{2}T\nu_{0}}{\pi}\frac{\Delta}{\omega_{\nu}+\Delta
}\sum_{|k|=0}^{\nu-1}L_0\left(\Omega_{k}\right)\left[\psi\left(\frac{1}{2}+\frac{2\omega_{\nu}-|\Omega_{k}|+\Delta}{4\pi T }\right)-\psi\left(\frac{1}{2}+\frac{|\Omega_{k}|+\Delta}{4\pi T}\right)\right]
\]
First of all one can easily see that the contributions of the positive
and negative $k$ are equal. The method to continue such sum on real
frequencies was developed in Ref. \onlinecite{AV80} and consists in Eliashberg
transformation (\ref{Eliashberg}) of the sum over $\Omega_{k}$ to integral over the
contour C (see Fig. \ref{phasediagram} ) (see the detailed
description of this procedure in Ref. \onlinecite{LV09}). Replacing $k\rightarrow-iz$
one finds
\begin{align*}
Q_{xx}^{\mathrm{MT(an)}}(\omega_{\nu}) & =-\frac{2e^{2}T\nu_{0}}{\pi}\frac{\Delta}{\omega_{\nu}+\Delta}\left\{ \frac{1}{2}L_0(0)\left[\psi\left(\frac{1}{2}+\frac{2\omega_{\nu}+\Delta}{4\pi T}\right)-\psi\left(\frac{1}{2}+\frac{\Delta}{4\pi T}\right)\right]\right.\\
 & +\left.\frac{1}{2i}{\displaystyle \oint_{C_{2}}}dz\coth\left(\pi z\right)L_0(-iz)\left[\psi\left(\frac{1}{2}+\frac{2\omega_{\nu}+\Delta}{4\pi T}+\frac{iz}{2}\right)-\psi\left(\frac{1}{2}-\frac{iz}{2}+\frac{\Delta}{4\pi T}\right)\right]\right\} .
\end{align*}

One can see that the integral over small semicircle around the point
$z=0$ compensates exactly the first term in the curly brackets. What
concerns the semicircle around the point $z=i\nu$ here the integrand
function is equal to zero. Hence we get:
\[
Q_{xx}^{\mathrm{MT(an)}}(\omega_{\nu})=2e^{2}T\frac{\Delta}{\omega_{\nu}+\Delta}\frac{1}{2\pi i}\left(\int_{-\infty}^{\infty}-\int_{-\infty+i\nu}^{\infty+i\nu}\right)dz\frac{\coth\left(\pi z\right)}{\widetilde{h}-\frac{2\pi Tiz}{\Delta}}\left[\psi\left(\frac{1}{2}+\frac{2\omega_{\nu}+\Delta}{4\pi T}+\frac{iz}{2}\right)-\psi\left(\frac{1}{2}+\frac{\Delta}{4\pi T}-\frac{iz}{2}\right)\right]
\]
Shifting the variables in the second integral over the upper line
$\Im z=\nu$, one can get the analytical expression valid for Matsubara
frequencies. Then one performs analytical continuation $\omega_{\nu}\rightarrow-i\omega$
and some simple variable shifting to get
\[
Q_{xx}^{\mathrm{MT(an)R}}(\omega)=-i\frac{e^{2}}{\pi^{2}}\frac{T}{1-i\omega/\Delta}\int_{-\infty}^{\infty}dz\left[\frac{\coth\left(z\right)-\coth\left(z-\frac{\omega}{2T}\right)}{\widetilde{h}-\frac{2Tiz}{\Delta}}\right]\left[\psi\left(\frac{1}{2}+\frac{-2i\omega+\Delta}{4\pi T}+\frac{iz}{2\pi}\right)-\psi\left(\frac{1}{2}+\frac{\Delta}{4\pi T}-\frac{iz}{2\pi}\right)\right]
\]

One can expand the $\psi$-functions by means of Eq. (\ref{psi_half}) in view of $T,\omega\ll\Delta$, and get
\[
Q_{xx}^{\mathrm{MT(an)R}}(\omega)\approx\frac{4e^{2}}{\pi^{2}\Delta}\frac{1+\frac{i\omega}{\Delta}}{1-\frac{i\omega}{\Delta}}T^{2}\int_{-\infty}^{\infty}dz\left[\frac{\coth\left(z\right)-\coth\left(z-\frac{\omega}{2T}\right)}{\widetilde{h}-\frac{2Tiz}{\Delta}}\right]\left(z-\frac{\omega}{2T}\right).
\]
The integration can be performed analytically; thus one arrives at
the final result:
\begin{equation}
\sigma_{xx}^{\mathrm{MT(an)}}(\omega)=\frac{2e^{2}}{\pi^{2}}\frac{1+\frac{i\omega}{\Delta}}{1-\frac{i\omega}{\Delta}}\left[\frac{2}{3}\left(\frac{\gamma_{E}t}{\widetilde{h}}\right)^{2}+\frac{\Delta\widetilde{h}}{i\omega}\left(1-\frac{i\omega}{\Delta\widetilde{h}}\right)\ln\left(1-\frac{i\omega}{\Delta\widetilde{h}}\right)+1\right]. 
\label{MTan}
\end{equation}

\[
\sigma_{xx}^{\mathrm{MT(an)}}(\omega\ll\Delta\widetilde{h})=\frac{e^{2}}{\pi^{2}}\left[\frac{i\omega}{\Delta\widetilde{h}}-\frac{1}{3}\left(\frac{\omega}{\Delta\widetilde{h}}\right)^{2}+\frac{4}{3}\left(\frac{\gamma_{E}t}{\widetilde{h}}\right)^{2}\right].
\]

\subsection{Regular part of the MT contribution}

In the LLL approximation the MT regular part (\ref{d21-1}) acquires the form 
\begin{equation}
Q_{xx}^{\mathrm{MT(reg1)}}(\omega_{\nu})=2e^{2}T\frac{1}{\omega_{\nu}}\frac{\Delta}{2\pi}\sum_{\Omega_{k}}\frac{1}{\tilde{h}+\frac{|\Omega_{k}|}{\Delta}}\left[\psi\left(\frac{1}{2}+\frac{2\omega_{\nu}+|\Omega_{k}|+\Delta}{4\pi T}\right)-\psi\left(\frac{1}{2}+\frac{|\Omega_{k}|+\Delta}{4\pi T}\right)\right].\label{d21-1-1}
\end{equation}
Now, as above was done, let us use the low temperature asymptotic of the $\psi$-functions Eq. (\ref{psi_half})
\begin{equation}
Q_{xx}^{\mathrm{MT(reg1)}}(\omega_{\nu})=2e^{2}T\frac{1}{\omega_{\nu}}\frac{\Delta}{2\pi}\sum_{\Omega_{k}}\frac{1}{\tilde{h}+\frac{|\Omega_{k}|}{\Delta}}\ln\left(1+\frac{2\omega_{\nu}}{|\Omega_{k}|+\Delta}\right). \label{d21-1-1-1}
\end{equation}
One may omit $\Omega_{k}$ in the logarithm since the singular contribution comes from the propagator, where $\tilde{h}\ll1$.

In purpose to obtain the frequency and the lowest-order temperature dependence, one can present the $k$-summation from $(-\infty,\infty)$ to $[0,\frac{\Delta}{2\pi T}\rightarrow\infty)$ in Eq. (\ref{d21-1-1-1}) as 
\begin{equation}
Q_{xx}^{\mathrm{MT(reg1)}}(\omega_{\nu})=2e^{2}T\frac{1}{\omega_{\nu}}\frac{\Delta}{\pi}\sum_{k=0}^{\frac{\Delta}{2\pi T}}\frac{1}{\tilde{h}+\frac{\Omega_{k}}{\Delta}}\ln\left(1+\frac{2\omega_{\nu}}{\Delta}\right)-2e^{2}T\frac{1}{\omega_{\nu}}\frac{\Delta}{2\pi}\frac{1}{\tilde{h}}\ln\left(1+\frac{2\omega_{\nu}}{\Delta}\right). \label{d21-1-1-1-2}
\end{equation}
The summation can be evaluated using the definition of $\psi$-function Eq. (\ref{psi_def}).  Using the low temperature asymptotic of the $\psi$-function Eq. (\ref{psi}) and relation (\ref{Delta_Tc}), one gets
\begin{equation*}
Q_{xx}^{\mathrm{MT(reg1)}}(\omega_{\nu}) =\frac{e^{2}}{\pi^{2}}\frac{\Delta^{2}}{\omega_{\nu}}\ln\left(1+\frac{2\omega_{\nu}}{\Delta}\right)\left[\ln\left(\frac{1}{\widetilde{h}}\right)+\frac{1}{3}\left(\frac{\gamma_{E}t}{\widetilde{h}}\right)^{2}\right]. 
\end{equation*}
Performing the analytical continuation $\omega_{\nu}\rightarrow-i\omega$, one finds 
\begin{equation}
\begin{split}
\sigma_{xx}^{\mathrm{MT(reg1)}}(\omega)	& =\frac{Q_{xx}^{\mathrm{MT(reg1)}}(\omega)-Q_{xx}^{\mathrm{MT(reg1)}}(0)}{-i\omega} \\
	 & =-\frac{e^{2}}{\pi^{2}}\left[\left(\frac{\Delta}{\omega}\right)^{2}\ln\left(1-\frac{2i\omega}{\Delta}\right)+\frac{2i\Delta}{\omega}\right]\left[\ln\left(\frac{1}{\tilde{h}}\right)+\frac{1}{3}\left(\frac{\gamma_{E}t}{\widetilde{h}}\right)^{2}\right]. 
\end{split}
\label{MTreg}
\end{equation}
Expanding the logarithm one finds the low frequency asymptotics 
\begin{equation}
\sigma_{xx}^{\mathrm{MT(reg1)}}(\omega\ll\Delta)=-\frac{2e^{2}}{\pi^{2}}\left(1+\frac{4i\omega}{3\Delta}-\frac{2\omega^2}{\Delta^2}\right)\left[\ln\left(\frac{1}{\tilde{h}}\right)+\frac{1}{3}\left(\frac{\gamma_{E}t}{\widetilde{h}}\right)^{2}\right].
\end{equation}

One can see that the contribution to the impedance (\ref{MTreg}) shows up itself at the frequencies in the scale $\omega\sim\Delta$, i.e. much later than those ones of AL and anomalous MT contributions.

\subsection{Contribution of the diagrams 3-4}

We now pass to the calculation of diagrams 3 and 4, which identify the renormalization of the single-particle diffusion coefficient in the presence of fluctuations. They give similar contributions as the regular part of the MT diagram.
\[
Q_{xx}^{(3)}(\omega_{\nu})=2e^{2}T^{2}\sum_{k,n}\int\frac{d^{2}\mathbf{q}}{(2\pi)^{2}}L\left(\mathbf{q},\Omega_{k}\right)\lambda\left(\mathbf{q},\varepsilon_{n},\Omega_{k}-\varepsilon_{n}\right)\lambda\left(\mathbf{q},\varepsilon_{n+\nu},\Omega_{k}-\varepsilon_{n+\nu}\right)C\left(\mathbf{q},\varepsilon_{n+\nu},\Omega_{k}-\varepsilon_{n}\right)I_{1}^{(3)}I_{2}^{(3)},
\]
where the integrals of the Green’s function products can be calculated
in the standard way:
\begin{align*}
I_{1}^{(3)}\left(\mathbf{q},\varepsilon_{n},\varepsilon_{n+\nu},\Omega_{k}-\varepsilon_{n}\right) & =\int\frac{d^{2}\mathbf{p}}{(2\pi)^{2}}v_{x}(\mathbf{p})G(\mathbf{p},\varepsilon_{n})G(\mathbf{p},\varepsilon_{n+\nu})G(\mathbf{q}-\mathbf{p},\Omega_{k}-\varepsilon_{n})\\
 & =-4\pi\nu_{0}\mathcal{D}\tau^{2}q_{x}\Theta\left(\varepsilon_{n}\varepsilon_{n+\nu}\right)\Theta\left(-\varepsilon_{n}\Omega_{k-n}\right)
\end{align*}

\begin{align*}
I_{2}^{(3)}\left(\mathbf{q},\Omega_{k}-\varepsilon_{n},\Omega_{k}-\varepsilon_{n+\nu},\varepsilon_{n+\nu}\right) & =-I_{1}^{(3)}\left(\mathbf{q},\Omega_{k}-\varepsilon_{n},\Omega_{k}-\varepsilon_{n+\nu},\varepsilon_{n+\nu}\right).
\end{align*}
Thus,
\[
Q_{xx}^{(3)}(\omega_{\nu})=-16\pi e^{2}T^{2}\nu_{0}\mathcal{D}\int\frac{d^{2}\mathbf{q}\left(\mathcal{D}q_{x}^{2}\right)}{(2\pi)^{2}}\sum_{k}L\left(\mathbf{q},\Omega_{k}\right)\sum_{n}\frac{\Theta\left(-\varepsilon_{n}\Omega_{k-n}\right)}{|2\varepsilon_{n}-\Omega_{k}|+\mathcal{D}q^{2}}\frac{\Theta\left(-\varepsilon_{n+\nu}\Omega_{k-n-\nu}\right)}{|2\varepsilon_{n}+2\omega_{\nu}-\Omega_{k}|+\mathcal{D}q^{2}}\frac{\Theta\left(\varepsilon_{n}\varepsilon_{n+\nu}\right)}{|2\varepsilon_{n}+\omega_{\nu}-\Omega_{k}|+\mathcal{D}q^{2}}
\]
And the summation of $n$ can be examined as 
\begin{align*}
\Sigma^{(3)} & =\sum_{n=-\infty}^{\infty}\frac{\Theta\left(-\varepsilon_{n}\Omega_{k-n}\right)}{|2\varepsilon_{n}-\Omega_{k}|+\mathcal{D}q^{2}}\frac{\Theta\left(-\varepsilon_{n+\nu}\Omega_{k-n-\nu}\right)}{|2\varepsilon_{n}+2\omega_{\nu}-\Omega_{k}|+\mathcal{D}q^{2}}\frac{\Theta\left(\varepsilon_{n}\varepsilon_{n+\nu}\right)}{|2\varepsilon_{n}+\omega_{\nu}-\Omega_{k}|+\mathcal{D}q^{2}}\\
 & =\frac{-1}{4\pi T\omega_{\nu}^{2}}\left[\psi\left(\frac{1}{2}+\frac{\left|\Omega_{k}\right|+\mathcal{D}q^{2}}{4\pi T}\right)+\psi\left(\frac{1}{2}+\frac{2\omega_{\nu}+\left|\Omega_{k}\right|+\mathcal{D}q^{2}}{4\pi T}\right)-2\psi\left(\frac{1}{2}+\frac{\omega_{\nu}+\left|\Omega_{k}\right|+\mathcal{D}q^{2}}{4\pi T}\right)\right].
\end{align*}

Let us now pass to the Landau representation, restricting by the LLL.

\begin{equation}
Q_{xx}^{(3-4)}(\omega_{\nu})=-16\pi e^{2}T^{2}\nu_{0}\frac{\Delta^2}{2\pi}\sum_{k}L_{0}\left(\Omega_{k}\right)\Sigma_{0}^{(3)}(\omega_{\nu},\Omega_{k}),
\label{Q7-10}
\end{equation}
with
\[
\Sigma_{0}^{(3)}(\omega_{\nu},\Omega_{k})=\frac{-1}{4\pi T\omega_{\nu}^{2}}\left[\psi\left(\frac{1}{2}+\frac{\left|\Omega_{k}\right|+\Delta}{4\pi T}\right)+\psi\left(\frac{1}{2}+\frac{2\omega_{\nu}+\left|\Omega_{k}\right|+\Delta}{4\pi T}\right)-2\psi\left(\frac{1}{2}+\frac{\omega_{\nu}+\left|\Omega_{k}\right|+\Delta}{4\pi T}\right)\right].
\]
One can do similar tricks as in dealing with Eq. (\ref{d21-1-1-1-2}) in regular MT subsection. Considering the singular contribution comes from the propagator $L_0(\Omega_k)$, one may omit $\Omega_{k}$ in the $\psi$-functions. Then the $k$-summation can be performed using the definition of $\psi$-function Eq. (\ref{psi_def}). And again using the low temperature asymptotic of the $\psi$-functions Eqs. (\ref{psi}) and (\ref{psi1}), after analytical continuation $\omega_{\nu}\rightarrow-i\omega$, the result reads 
\begin{equation*}
Q_{xx}^{(3-4)}(\omega)=\frac{2e^{2}\Delta^{3}}{\pi^2\omega^{2}}\ln\left(\frac{1-\frac{2i\omega}{\Delta}}{\left(1-\frac{i\omega}{\Delta}\right)^{2}}\right)\left[\ln\left(\frac{1}{\widetilde{h}}\right)+\frac{1}{3}\left(\frac{\gamma_{E}t}{\widetilde{h}}\right)^{2}\right]. 
\end{equation*}
One can see that the corresponding contribution to the impedance, similarly to the situation with the regular MT one,  appears at the frequencies of the external electromagnetic field in the scale $\omega\sim\Delta$, i.e. much later than in AL and anomalous MT  contributions.
Finally, one obtains
\begin{align*}
\sigma_{xx}^{\mathrm{MT}(3-4)}(\omega) =\frac{2e^{2}}{\pi^{2}}\left[\ln\frac{1}{\widetilde{h}}+\frac{1}{3}\left(\frac{\gamma_{E}t}{\widetilde{h}}\right)^{2}\right] \left[\left(\frac{\Delta}{i\omega}\right)^{3}\ln\frac{1-\frac{2i\omega}{\Delta}}{\left(1-\frac{i\omega}{\Delta}\right)^{2}}+\frac{\Delta}{i\omega}\right] 
\end{align*}
Expanding the logarithm one finds the low frequency asymptotics 
\begin{equation}
\sigma_{xx}^{\mathrm{MT}(3-4)}(\omega\ll\Delta)=-\frac{4e^{2}}{\pi^{2}}\left[\ln\left(\frac{1}{\widetilde{h}}\right)+\frac{1}{3}\left(\frac{\gamma_{E}t}{\widetilde{h}}\right)^{2}\right]\left(1+\frac{7}{4}\frac{i\omega}{\Delta}-\frac{3\omega^{2}}{\Delta^{2}}\right).
\end{equation}

\section{DOS (diagrams 5-8) Contribution and their modifications (diagrams 9-10)}
\subsection{Basic Expressions}

As above we use the intermediate results of \cite{SM-LV09} for the diagrams
and then quantize the motion of the center of mass of the Cooper pair
in a magnetic field. 

\subsubsection{Diagram 5}
The general expression for the diagram 5 is read
as
\begin{equation}
Q_{xx}^{(5)}(\omega_{\nu})=2e^{2}T^{2}\int\frac{d^{2}q}{(2\pi)^{2}}\sum_{k,n}L\left(q,\Omega_{k}\right)\left[\lambda^{2}\left(q,\varepsilon_{n},\Omega_{k}-\varepsilon_{n}\right)\right]I_{xx}^{(5)}, \label{q5}
\end{equation}
with the integral $I_{xx}^{(5)}$ of four electron Green functions
calculated exactly in \cite{LV09} in the same spirit as it was demonstrated
above: 
\begin{align*}
I_{xx}^{(5)} & =\int\frac{d^{2}p}{(2\pi)^{2}}v_{x}^{2}~G^{2}\left(p,\varepsilon_{n}\right)G\left(p,\varepsilon_{n}+\omega_{\nu}\right)G\left(p,\Omega_{k}-\varepsilon_{n}\right)\\
& =\nu_{0}\mathcal{D}\tau^{-1}\int d\xi\frac{1}{\left(\xi-i\tilde{\varepsilon}_{n}\right)^{2}}\frac{1}{\left(\xi-i\tilde{\varepsilon}_{n+\nu}\right)}\frac{1}{\left(\xi+i\tilde{\varepsilon}_{n-k}\right)}.
\end{align*}

Taking into account all possible positions of the poles one can write
\begin{align*}
I_{xx}^{(5)} & =2\pi i\nu_{0}\mathcal{D}\tau^{-1}\Theta\left(-{\varepsilon}_{n+\nu}{\varepsilon}_{n}\right)\left[\frac{\Theta\left({\varepsilon}_{n-k}{\varepsilon}_{n}\right)}{\left(i\tilde{\varepsilon}_{n+\nu}-i\tilde{\varepsilon}_{n}\right)^{2}\left(i\tilde{\varepsilon}_{n+\nu}+i\tilde{\varepsilon}_{n-k}\right)}-\frac{\Theta\left({\varepsilon}_{n-k}{\varepsilon}_{n}\right)}{\left(i\tilde{\varepsilon}_{n-k}+i\tilde{\varepsilon}_{n}\right)^{2}\left(i\tilde{\varepsilon}_{n-k}+i\tilde{\varepsilon}_{n+\nu}\right)}\right]\\
 & +2\pi i\nu_{0}\mathcal{D}\tau^{-1}\Theta\left(-{\varepsilon}_{n+\nu}{\varepsilon}_{n}\right)\frac{\Theta\left(-{\varepsilon}_{n-k}{\varepsilon}_{n}\right)}{\left(i\tilde{\varepsilon}_{n+\nu}-i\tilde{\varepsilon}_{n}\right)^{2}\left(i\tilde{\varepsilon}_{n+\nu}+i\tilde{\varepsilon}_{n-k}\right)}+  2\pi i\nu_{0}\mathcal{D}\tau^{-1}\frac{\Theta\left({\varepsilon}_{n+\nu}{\varepsilon}_{n}\right)\Theta\left({\varepsilon}_{n-k}{\varepsilon}_{n}\right)\mathrm{sgn}\left({\varepsilon}_{n-k}\right)}{\left(-i\tilde{\varepsilon}_{n-k}-i\tilde{\varepsilon}_{n}\right)^{2}\left(i\tilde{\varepsilon}_{n+\nu}+i\tilde{\varepsilon}_{n-k}\right)}.
\end{align*}

We arrive to
\[
I_{xx}^{(5)}\approx-2\pi\nu_0\mathcal{D}\tau^{2}\left[\Theta\left(-{\varepsilon}_{n+\nu}{\varepsilon}_{n}\right)\Theta\left(-{\varepsilon}_{n-k}{\varepsilon}_{n}\right)+\Theta\left({\varepsilon}_{n+\nu}{\varepsilon}_{n}\right)\Theta\left({\varepsilon}_{n-k}{\varepsilon}_{n}\right)-2\Theta\left(-{\varepsilon}_{n+\nu}{\varepsilon}_{n}\right)\Theta\left({\varepsilon}_{n-k}{\varepsilon}_{n}\right)\right].
\]

We have to take into account that the vertex $\lambda\left(q,\varepsilon_{n},\Omega_{k}-\varepsilon_{n}\right)$
contains the theta function $\Theta\left({\varepsilon}_{n-k}{\varepsilon}_{n}\right)$. Hence,
in the following we can use the reduced piece of $I_{xx}^{(5)}$:
\[
\Delta I_{xx}^{(5)}\approx-2\pi\nu_0\mathcal{D}\tau^{2}\Theta\left({\varepsilon}_{n-k}{\varepsilon}_{n}\right)\left[\Theta\left({\varepsilon}_{n+\nu}{\varepsilon}_{n}\right)-2\Theta\left(-{\varepsilon}_{n+\nu}{\varepsilon}_{n}\right)\right].
\]

\subsubsection{Diagram 7}
Let us proceed to the discussion of diagram 7. Its contribution can be
written in the same way as above:
\begin{equation*}
Q_{xx}^{\left(7\right)}(\omega_{\nu})=2e^{2}T^{2}\int\frac{d^{2}q}{(2\pi)^{2}}\sum_{n,k}L\left(q,\Omega_{k}\right)\left[\lambda^{2}\left(q,\varepsilon_{n},\Omega_{k}-\varepsilon_{n}\right)\right]\frac{v_{F}^{2}}{2}\left(\frac{1}{2\pi\nu_{0}\tau}\right)I_{1}I_{2}, \label{q7v}
\end{equation*}
where the integrals are
\begin{equation*}
I_{1}=\int\frac{d^{2}p}{(2\pi)^{2}}~G^{2}\left(p,\varepsilon_{n}\right)G\left(p,\varepsilon_{n}+\omega_{\nu}\right)=2\pi i\nu_{0}\tau^{2}\mathrm{sgn}\left(\varepsilon_{n+\nu}\right)\Theta\left(-\varepsilon_{n+\nu}\varepsilon_{n}\right),
\end{equation*}
and
\begin{equation*}
I_{2}=\int\frac{d^{2}p^{\prime}}{(2\pi)^{2}}~G^{2}\left(p^{\prime},\varepsilon_{n}\right)G\left(q-p^{\prime},\Omega_{k}-\varepsilon_{n}\right)=-2\pi i\nu_{0}\tau^{2}\mathrm{sgn}\left(\varepsilon_{n}\right)\Theta\left(\varepsilon_{n-k}\varepsilon_{n}\right).
\end{equation*}
The product is
\begin{align*}
\frac{v_{F}^{2}}{2}\left(\frac{1}{2\pi\nu_{0}\tau}\right)I_{1}I_{2} & =-\frac{v_{F}^{2}}{2}\left(\frac{1}{2\pi\nu_{0}\tau}\right)4\pi^{2}\nu_{0}^{2}\tau^{4}\Theta\left(-\varepsilon_{n+\nu}\varepsilon_{n}\right)\Theta\left(\varepsilon_{n-k}\varepsilon_{n}\right)\label{q7v-1}\\
&=-2\pi\mathcal{D}\nu_{0}\tau^{2}\Theta\left(-\varepsilon_{n+\nu}\varepsilon_{n}\right)\Theta\left(\varepsilon_{n-k}\varepsilon_{n}\right).
\end{align*}

\subsubsection{Sum of the diagrams 5 and 7}
Now let us sum two contributions:
\begin{equation*}
Q_{xx}^{(5+7)}(\omega_{\nu})=2e^{2}T^{2}\int\frac{d^{2}q}{(2\pi)^{2}}\sum_{k,n}L\left(q,\Omega_{k}\right)\left[\lambda^{2}\left(q,\varepsilon_{n},\Omega_{k}-\varepsilon_{n}\right)\right]\left(\Delta I_{xx}^{(5)}+\frac{v_{F}^{2}}{2}\left(\frac{1}{2\pi\nu_{0}\tau}\right)I_{1}I_{2}\right),
\end{equation*}
where
\begin{align*}
\Delta I_{xx}^{(5)}+\frac{v_{F}^{2}}{2}\left(\frac{1}{2\pi\nu_{0}\tau}\right)I_{1}I_{2} =-2\pi\nu_{0}\mathcal{D}\tau^{2}\Theta\left({\varepsilon}_{n-k}{\varepsilon}_{n}\right)\mathrm{sgn}\left({\varepsilon}_{n+\nu}{\varepsilon}_{n}\right).
\end{align*}
Thus
\begin{align*}
Q_{xx}^{(5+7)}(\omega_{\nu}) & =-4\pi\nu_{0}\mathcal{D}e^{2}T^{2}\int\frac{d^{2}q}{(2\pi)^{2}}\sum_{k=-\infty}^{\infty}L\left(q,\left|\Omega_{k}\right|\right)\sum_{n=-\infty}^{\infty}\frac{\Theta\left({\varepsilon}_{n-k}{\varepsilon}_{n}\right)\mathrm{sgn}\left({\varepsilon}_{n+\nu}{\varepsilon}_{n}\right)}{\left(|2\varepsilon_{n}-\Omega_{k}|+\mathcal{D}q^{2}\right)^{2}} \label{q7h-1}\\
= & -4\pi\nu_{0}\mathcal{D}e^{2}T^{2}\int\frac{d^{2}q}{(2\pi)^{2}}\sum_{k=-\infty}^{\infty}L\left(q,\left|\Omega_{k}\right|\right)\left[\sum_{n=-\infty}^{\infty}\frac{\Theta\left(\varepsilon_{n}\varepsilon_{n-k}\right)}{\left(\left\vert 2\varepsilon_{n}-\Omega_{k}\right\vert +\mathcal{D}q^{2}\right)^{2}}-2\sum_{n=-\nu}^{-1}\frac{\Theta\left(\Omega_{k}-\varepsilon_{n}\right)}{\left(\left\vert 2\varepsilon_{n}-\Omega_{k}\right\vert +\mathcal{D}q^{2}\right)^{2}}\right].
\end{align*}
The first sum is independent of the external frequency $\omega_{\nu}$
and is consequently canceled out by analogous contributions from the
remaining diagrams. 

Let us calculate the second sum and start from the replacement $n\rightarrow-n$:
\begin{align*}
\widetilde{Q}_{xx}^{(5+7)}(\omega_{\nu}) &  =8\pi\nu_{0}\mathcal{D}e^{2}T^{2}\int\frac{d^{2}q}{(2\pi)^{2}}\sum_{k=-\infty}^{\infty}L\left(q,\left|\Omega_{k}\right|\right)\sum_{n=0}^{\nu-1}\frac{\Theta\left(\varepsilon_{n+k}\right)}{\left(\left\vert 2\varepsilon_{n+k}-\Omega_{k}\right\vert +\mathcal{D}q^{2}\right)^{2}}\\
 & =8\pi\nu_{0}\mathcal{D}e^{2}T^{2}\int\frac{d^{2}q}{(2\pi)^{2}}\left[\sum_{k=1}^{\nu-1}L\left(q,\left|\Omega_{k}\right|\right)\sum_{n=0}^{\nu-k-1}\frac{1}{\left(2\varepsilon_{n}+\Omega_{k}+\mathcal{D}q^{2}\right)^{2}}+\sum_{k=0}^{\infty}L\left(q,\left|\Omega_{k}\right|\right)\sum_{n=0}^{\nu-1}\frac{1}{\left(2\varepsilon_{n}+\Omega_{k}+\mathcal{D}q^{2}\right)^{2}}\right],
\end{align*}
which leads to
\begin{align*}
\widetilde{Q}_{xx}^{(5+7)}(\omega_{\nu}) = & \frac{\nu_0\mathcal{D}e^{2}}{2\pi}\int\frac{d^{2}q}{(2\pi)^{2}}\sum_{k=1}^{\nu-1}L\left(q,\left|\Omega_{k}\right|\right)\left[\psi'\left(\frac{1}{2}+\frac{\Omega_{k}+\mathcal{D}q^{2}}{4\pi T}\right)-\psi'\left(\frac{1}{2}+\frac{2\omega_{\nu}-\Omega_{k}+\mathcal{D}q^{2}}{4\pi T}\right)\right]\\
& + \frac{\nu_0\mathcal{D}e^{2}}{2\pi}\int\frac{d^{2}q}{(2\pi)^{2}}\sum_{k=0}^{\infty}L\left(q,\left|\Omega_{k}\right|\right)\left[\psi'\left(\frac{1}{2}+\frac{\Omega_{k}+\mathcal{D}q^{2}}{4\pi T}\right)-\psi'\left(\frac{1}{2}+\frac{2\omega_{\nu}+\Omega_{k}+\mathcal{D}q^{2}}{4\pi T}\right)\right].
\end{align*}

\subsubsection{Diagram 9}
Let us proceed to the similar contributions from diagrams 9-10 that consider the renormalized one-electron diffusion coefficient in the presence of fluctuation pairing.
\[
Q_{xx}^{(9)}(\omega_{\nu})=2e^{2}T^{2}\sum_{k,n}\int\frac{d^{2}\mathbf{q}}{(2\pi)^{2}}L\left(\mathbf{q},\Omega_{k}\right)\left[\lambda\left(\mathbf{q},\varepsilon_{n},\Omega_{k}-\varepsilon_{n}\right)\right]^{2}C\left(\mathbf{q},\varepsilon_{n+\nu},\Omega_{k}-\varepsilon_{n}\right)\left[I_{1}^{(9)}\right]^{2},
\]
where the integrals of the Green’s function products can be calculated
as in diagram 3:
\begin{align*}
I_{1}^{(9)}\left(\mathbf{q},\varepsilon_{n},\varepsilon_{n+\nu},\Omega_{k}-\varepsilon_{n}\right) & =I_{1}^{(3)}\left(\mathbf{q},\varepsilon_{n},\varepsilon_{n+\nu},\Omega_{k}-\varepsilon_{n}\right)
\end{align*}
Thus,
\begin{equation}
Q_{xx}^{(9)}(\omega_{\nu})=16\pi e^{2}T^{2}\nu_{0}\mathcal{D}\int\frac{d^{2}\mathbf{q}\left(\mathcal{D}q_{x}^{2}\right)}{(2\pi)^{2}}\sum_{k}L\left(\mathbf{q},\Omega_{k}\right)\sum_{n}\frac{\Theta\left(-\varepsilon_{n}\Omega_{k-n}\right)}{\left(|2\varepsilon_{n}-\Omega_{k}|+\mathcal{D}q^{2}\right)^{2}}\frac{\Theta\left(\varepsilon_{n}\varepsilon_{n+\nu}\right)}{|2\varepsilon_{n}+\omega_{\nu}-\Omega_{k}|+\mathcal{D}q^{2}}
\label{Q_9}
\end{equation}
The summation of $n$ can be examined in two parts 
\begin{align*}
\Sigma^{(9)} & =\sum_{n=-\infty}^{\infty}\frac{\Theta\left(-\varepsilon_{n}\Omega_{k-n}\right)}{\left(|2\varepsilon_{n}-\Omega_{k}|+\mathcal{D}q^{2}\right)^{2}}\frac{\Theta\left(\varepsilon_{n}\varepsilon_{n+\nu}\right)}{|2\varepsilon_{n}+\omega_{\nu}-\Omega_{k}|+\mathcal{D}q^{2}} \equiv\Sigma_{1}^{(9)}+\Sigma_{2}^{(9)}, \\
 % & =\sum_{n=0}^{\infty}\frac{\Theta\left(\Omega_{k}+\varepsilon_{n}+\omega_{\nu}\right)}{\left(|2\varepsilon_{n}+2\omega_{\nu}+\Omega_{k}|+\mathcal{D}q^{2}\right)^{2}\left(|2\varepsilon_{n}+\Omega_{k}+\omega_{\nu}|+\mathcal{D}q^{2}\right)}+\sum_{n=0}^{\infty}\frac{\Theta\left(\varepsilon_{n}+\Omega_{k}\right)}{\left(|2\varepsilon_{n}+\Omega_{k}|+\mathcal{D}q^{2}\right)^{2}\left(|2\varepsilon_{n}+\omega_{\nu}+\Omega_{k}|+\mathcal{D}q^{2}\right)}
\end{align*}
with
\begin{align*}
\Sigma_{1}^{(9)} = &-\sum_{n=0}^{\nu-1}\frac{\Theta(\varepsilon_{n}-\Omega_{k})}{\left(2\varepsilon_{n}-\Omega_{k}+\mathcal{D}q^{2}\right)^{2}\left(2\varepsilon_{n}-\omega_{\nu}-\Omega_{k}+\mathcal{D}q^{2}\right)}\\
= & -\frac{\Theta(\omega_{\nu}-\Omega_{k})}{4\pi T\omega_{\nu}^{2}}\left[-\psi\left(\frac{1}{2}+\frac{-\omega_{\nu}+\left|\Omega_{k}\right|+\mathcal{D}q^{2}}{4\pi T}\right)+\psi\left(\frac{1}{2}+\frac{\left|\Omega_{k}\right|+\mathcal{D}q^{2}}{4\pi T}\right)-\frac{\omega_{\nu}}{4\pi T}\psi'\left(\frac{1}{2}+\frac{\left|\Omega_{k}\right|+\mathcal{D}q^{2}}{4\pi T}\right)\right]\\
& -\frac{\Theta(\omega_{\nu}-\Omega_{k})}{4\pi T\omega_{\nu}^{2}}\left[\psi\left(\frac{1}{2}+\frac{\omega_{\nu}+\left|\Omega_{k}\right|+\mathcal{D}q^{2}}{4\pi T}\right)-\psi\left(\frac{1}{2}+\frac{2\omega_{\nu}+\left|\Omega_{k}\right|+\mathcal{D}q^{2}}{4\pi T}\right)+\frac{\omega_{\nu}}{4\pi T}\psi'\left(\frac{1}{2}+\frac{2\omega_{\nu}+\left|\Omega_{k}\right|+\mathcal{D}q^{2}}{4\pi T}\right)\right],
\end{align*}
and 
\begin{align*}
\Sigma_{2}^{(9)} & =\sum_{n=0}^{\infty}\frac{\Theta(\varepsilon_{n}-\Omega_{k})}{\left(2\varepsilon_{n}-\Omega_{k}+\mathcal{D}q^{2}\right)^{2}\left(2\varepsilon_{n}-\omega_{\nu}-\Omega_{k}+\mathcal{D}q^{2}\right)}+\sum_{n=1}^{\infty}\frac{\Theta(\varepsilon_{n}+\Omega_{k})}{\left(2\varepsilon_{n}+\Omega_{k}+\mathcal{D}q^{2}\right)^{2}\left(2\varepsilon_{n}+\omega_{\nu}+\Omega_{k}+\mathcal{D}q^{2}\right)} \\
& =\frac{1}{4\pi T\omega_{\nu}^{2}}\left[-\psi\left(\frac{1}{2}+\frac{-\omega_{\nu}+|\Omega_{k}|+\mathcal{D}q^{2}}{4\pi T}\right)-\psi\left(\frac{1}{2}+\frac{\omega_{\nu}+|\Omega_{k}|+\mathcal{D}q^{2}}{4\pi T}\right)+2\psi\left(\frac{1}{2}+\frac{|\Omega_{k}|+\mathcal{D}q^{2}}{4\pi T}\right)\right].
\end{align*}

\subsection{DOS contribution close to $H_{c2}(0)$ (LLL approximation)}

\subsubsection{Diagrams 5-8}
Using the the formulated rules in Sec. \ref{MtoL}, one can pass from the momentum integration to  summation over the Landau levels. Accounting only for the lowest one, one gets 
\begin{equation}
\widetilde{Q}_{xx}^{(5 + 7)}({\omega _\nu })=\widetilde{Q}_{xx\left( 1 \right)}^{(5 + 7)}({\omega _\nu })+\widetilde{Q}_{xx\left( 2 \right)}^{(5 + 7)}({\omega _\nu }),
\end{equation}
where
\begin{align}
 \widetilde{Q}_{xx(1)}^{(5+7)}(\omega_{\nu}) = & -\frac{e^{2}\Delta}{4\pi^2}\sum_{k=0}^{\nu-1}\frac{1}{\widetilde{h}+\Omega_{k}/\Delta}\left[\psi'\left(\frac{1}{2}+\frac{\Omega_{k}+\Delta}{4\pi T}\right)-\psi'\left(\frac{1}{2}+\frac{2\omega_{\nu}-\Omega_{k}+\Delta}{4\pi T}\right)\right] \nonumber \\
 & +\frac{e^{2}\Delta}{8\pi^2}\frac{1}{\widetilde{h}}\left[\psi'\left(\frac{1}{2}+\frac{\Delta}{4\pi T}\right)-\psi'\left(\frac{1}{2}+\frac{2\omega_{\nu}+\Delta}{4\pi T}\right)\right],  \label{dos1}
 \end{align}
 and
 \begin{align}
 \widetilde{Q}_{xx(2)}^{(5+7)}(\omega_{\nu}) = & - \frac{e^{2}\Delta}{4\pi^2}\sum_{k=0}^{\infty}\frac{1}{\widetilde{h}+\Omega_{k}/\Delta}\left[\psi'\left(\frac{1}{2}+\frac{\Omega_{k}+\Delta}{4\pi T}\right)-\psi'\left(\frac{1}{2}+\frac{2\omega_{\nu}+\Omega_{k}+\Delta}{4\pi T}\right)\right] \nonumber\\
 & +\frac{e^{2}\Delta}{8\pi^2}\frac{1}{\widetilde{h}}\left[\psi'\left(\frac{1}{2}+\frac{\Delta}{4\pi T}\right)-\psi'\left(\frac{1}{2}+\frac{2\omega_{\nu}+\Delta}{4\pi T}\right)\right]. \label{dos2}
\end{align}

It can be seen that the $\widetilde{Q}_{xx\left(1\right)}^{(5 + 7)}$ and $\widetilde{Q}_{xx\left(2\right)}^{(5 + 7)}$ are respectively similar to the anomalous (\ref{d21-2}) and regular (\ref{d21-1}) contributions of the MT diagram. Therefore, we can handle them in a similar manner.

For the regular part $\widetilde{Q}_{xx\left( 2 \right)}^{(5 + 7)}$, since the singular part comes from the propagator, we can again neglect the $\Omega_k$ dependence of the $\psi'$-functions and use the asymptotic expansion Eq. (\ref{psi1_half}) of $\psi'(\frac{1}{2}+z)$ to $\frac{1}{z}$ term with the precision of $T/\Delta$. The result of Eq. (\ref{dos2}) after analytical continuation $\omega_{\nu}\rightarrow-i\omega$ can be found as  
\begin{align}
\widetilde{Q}_{xx(2)}^{(5+7)R}(\omega) =\frac{ e^{2}}{\pi^{2}}\frac{i\omega}{1-\frac{2i\omega}{\Delta}}\left[\ln\left(\frac{1}{\widetilde{h}}\right)+\frac{\pi^2}{3}\left(\frac{ T}{\Delta\widetilde{h}}\right)^{2}\right]. 
\label{QDOS1}
\end{align}

For the anomalous part $\widetilde{Q}_{xx\left(1\right)}^{(5 + 7)}$, one can again use the Eliashberg transformation (\ref{Eliashberg}) converting the sum over $k$ to integral over the
contour C in Fig. \ref{phasediagram}, as  
\begin{align}
    \widetilde{Q}_{xx(1)}^{(5+7)}(\omega_{\nu}) & = -\frac{e^{2}\Delta}{4\pi^{2}}\frac{1}{2i}\left(\int_{-\infty}^{\infty}-\int_{-\infty+i\nu}^{\infty+i\nu}\right)\frac{dz\coth\left(\pi z\right)}{\widetilde{h}-\frac{2\pi T}{\Delta}iz}\left[\psi'\left(\frac{1}{2}+\frac{\Delta}{4\pi T}-\frac{iz}{2}\right)-\psi'\left(\frac{1}{2}+\frac{2\omega_{\nu}+\Delta}{4\pi T}+\frac{iz}{2}\right)\right] \nonumber.
\label{QDOS2}
\end{align}
Shifting the variables in the second integral over the upper line
$\Im z=\nu$, one can get the analytical expression valid for Matsubara
frequencies. Then one performs analytical continuation $\omega_{\nu}\rightarrow-i\omega$
and some simple variable shifting to get
\begin{align*}
\widetilde{Q}_{xx(1)}^{(5+7)R}(\omega)=i\frac{e^{2}\Delta}{8\pi^{2}}\int_{-\infty}^{\infty}dz\left[\frac{\coth\left(\pi z\right)-\coth\left(\pi z-\frac{\omega}{2T}\right)}{\widetilde{h}-\frac{2\pi T}{\Delta}iz}\right]\left[\psi'\left(\frac{1}{2}+\frac{\Delta}{4\pi T}-\frac{iz}{2}\right)-\psi'\left(\frac{1}{2}+\frac{-2i\omega+\Delta}{4\pi T}+\frac{iz}{2}\right)\right]
\end{align*}
Expanding the $\psi'$-functions by (\ref{psi1_half}), the integration can be performed analytically as
\begin{align*}
 \widetilde{Q}_{xx(1)}^{(5+7)R}(\omega) = \frac{e^{2}}{\pi^{2}}\frac{-i\omega}{1-\frac{2i\omega}{\Delta}}\left\{\frac{\pi^2}{3}\left(\frac{ T}{\Delta\widetilde{h}}\right)^{2}\left[\left(\frac{1}{1-\frac{i\omega}{\Delta\widetilde{h}}}\right)^{2}-1\right]-\ln\left(\frac{1-\frac{i\omega}{\Delta\widetilde{h}}}{1-\frac{i\omega}{\Delta}}\right)\right\}.
\end{align*}

Taking into account the existence of other two equivalent diagrams (6 and 8), hence multiplying the result by 2, one can get the corresponding impedance 
\begin{align}
& \sigma_{xx(1)}^{\mathrm{DOS}(5-8)}(\omega)=-\frac{2e^{2}}{\pi^{2}}\frac{1}{1-\frac{2i\omega}{\Delta}}\left\{\frac{1}{3}\left(\frac{ \gamma_Et}{\widetilde{h}}\right)^{2}\left[1-\left(\frac{1}{1-\frac{i\omega}{\Delta\widetilde{h}}}\right)^{2}\right]+\ln\left(\frac{1-\frac{i\omega}{\Delta\widetilde{h}}}{1-\frac{i\omega}{\Delta}}\right)\right\}, \\
& \sigma_{xx(2)}^{\mathrm{DOS}(5-8)}(\omega)= -\frac{2e^{2}}{\pi^{2}}\frac{1}{1-\frac{2i\omega}{\Delta}}\left[\ln\left(\frac{1}{\widetilde{h}}\right)+\frac{1}{3}\left(\frac{\gamma_E t}{\widetilde{h}}\right)^{2}\right]. 
\end{align}
It is interesting to compare these contributions with the corresponding anomalous (\ref{MTan}) and regular (\ref{MTreg}) contributions of MT diagram. One can see that the $\sigma_{xx(2)}^{\mathrm{DOS}(5-8)}(\omega)$ and $\sigma_{xx}^{MT(reg1)}(\omega)$ show similar behaviors in the region of quantum fluctuations $t \ll \tilde h \ll 1$: they give singular contributions to fluctuation conductivity at static limit, while their frequency dependence show up in the scale $\omega \sim \Delta$, much later than the $\sigma_{xx(1)}^{\mathrm{DOS}(5-8)}(\omega)$ and $\sigma_{xx}^{MT(an)}(\omega)$ ones which appear in the scale $\omega \sim \Delta\widetilde{h}$.  However, the latter two only play minor roles in the static limit.

Finally, one finds the frequency dependent expression for the DOS contribution in the region of quantum fluctuations:
\begin{align}
\sigma_{xx}^{\mathrm{DOS}(5-8)}(\omega) = -\frac{2e^{2}}{\pi^{2}}\frac{1}{1-\frac{2i\omega}{\Delta}}\left\{\ln\left(\frac{1}{\widetilde{h}}\right)+\frac{1}{3}\left[2-\frac{1}{\left(1-\frac{i\omega}{\Delta\widetilde{h}}\right)^{2}}\right]\left(\frac{\gamma_{E}t}{\widetilde{h}}\right)^{2} -\ln\left(1-\frac{i\omega}{\Delta}\right)+\ln\left(1-\frac{i\omega}{\Delta\widetilde{h}}\right)\right\}, 
\label{DOSomega}
\end{align}
or, in the low frequency limit:
\begin{equation}
\sigma_{xx}^{\mathrm{DOS}(5-8)}(\omega\ll \Delta\widetilde{h}) = -\frac{2e^{2}}{\pi^{2}}\left[\ln\left(\frac{1}{\widetilde{h}}\right)+\frac{1}{3}\left(\frac{\gamma_{E}t}{\widetilde{h}}\right)^{2}-\frac{i\omega}{\Delta\widetilde{h}}+\frac{1}{2}\left(\frac{\omega}{\Delta\widetilde{h}}\right)^2\right].
\label{sgima58}
\end{equation}

\subsubsection{Diagrams 9-10}

Let us pass Eq. (\ref{Q_9}) to the Landau representation, restricting by the LLL
\begin{equation}
Q_{xx}^{(9-10)}(\omega_{\nu})=16\pi e^{2}T^{2}\nu_{0}\frac{\Delta^2}{2\pi}\sum_{k}L_{0}\left(\Omega_{k}\right)\left[\Sigma_{0(1)}^{(9)}(\omega_{\nu},\Omega_{k})+\Sigma_{0(2)}^{(9)}(\omega_{\nu},\Omega_{k})\right],
\label{Q7-10}
\end{equation}
with
\begin{align*}
\Sigma_{0(1)}^{(9)}(\omega_{\nu},\Omega_{k})= & -\frac{\Theta(\omega_{\nu}-\Omega_{k})}{4\pi T\omega_{\nu}^{2}}\left[-\psi\left(\frac{1}{2}+\frac{-\omega_{\nu}+\left|\Omega_{k}\right|+\Delta}{4\pi T}\right)+\psi\left(\frac{1}{2}+\frac{\left|\Omega_{k}\right|+\Delta}{4\pi T}\right)-\frac{\omega_{\nu}}{4\pi T}\psi'\left(\frac{1}{2}+\frac{\left|\Omega_{k}\right|+\Delta}{4\pi T}\right)\right] \\
	& -\frac{\Theta(\omega_{\nu}-\Omega_{k})}{4\pi T\omega_{\nu}^{2}}\left[\psi\left(\frac{1}{2}+\frac{\omega_{\nu}+\left|\Omega_{k}\right|+\Delta}{4\pi T}\right)-\psi\left(\frac{1}{2}+\frac{2\omega_{\nu}+\left|\Omega_{k}\right|+\Delta}{4\pi T}\right)+\frac{\omega_{\nu}}{4\pi T}\psi'\left(\frac{1}{2}+\frac{2\omega_{\nu}+\left|\Omega_{k}\right|+\Delta}{4\pi T}\right)\right],
\end{align*}
and
\[
\Sigma_{0(2)}^{(9)}(\omega_{\nu},\Omega_{k})=\frac{1}{4\pi T\omega_{\nu}^{2}}\left[-\psi\left(\frac{1}{2}+\frac{-\omega_{\nu}+|\Omega_{k}|+\Delta}{4\pi T}\right)-\psi\left(\frac{1}{2}+\frac{\omega_{\nu}+|\Omega_{k}|+\Delta}{4\pi T}\right)+2\psi\left(\frac{1}{2}+\frac{|\Omega_{k}|+\Delta}{4\pi T}\right)\right].
\]

One can evaluate Eq. (\ref{Q7-10}) separately, as
\begin{align*}
&Q_{xx(1)}^{(9-10)}(\omega_{\nu})  =8 e^{2}T^{2}\nu_{0}\Delta^2\sum_{k=-\infty}^{\infty} L_{0}\left(\Omega_{k}\right)\Sigma_{0(1)}^{(9)}(\omega_{\nu},\Omega_{k}) \\
= &-\frac{e^{2}\Delta}{\pi^{2}}\left\{\left(\frac{\Delta}{\omega_{\nu}}\right)^{2}\ln\left[\frac{\left(1-\frac{\omega_{\nu}}{\Delta}\right)\left(1+\frac{2\omega_{\nu}}{\Delta}\right)}{1+\frac{\omega_{\nu}}{\Delta}}\right]+\frac{2}{1+\frac{2\omega_{\nu}}{\Delta}}\right\}\left\{\ln\left(1+\frac{\omega_{\nu}}{\Delta\widetilde{h}}\right)+\ln\left(\frac{1}{\widetilde{h}}\right)+\frac{1}{3}\left(\frac{\gamma_{E}t}{\tilde{h}}\right)^{2}\left[2-\frac{1}{\left(1+\frac{\omega_{\nu}}{\Delta\widetilde{h}}\right)^{2}}\right]\right\}.
\end{align*}
and 
% Then using Eliashberg transformation (\ref{Eliashberg}) to convert the summation into integration, we have
\begin{equation*}
\begin{split}
Q_{xx(2)}^{(9-10)}(\omega_{\nu})& = 8e^{2}\nu_{0}T^{2}\Delta^{2}\sum_{k}L_{0}\left(\Omega_{k}\right)\Sigma_{0(2)}^{(9)}(\omega_{\nu},\Omega_{k}) \\
& =\frac{2e^{2}}{\pi^{2}}\frac{\Delta^{3}}{\omega_{\nu}^{2}}\ln\left(1-\frac{\omega_{\nu}^{2}}{\Delta^{2}}\right)\left[\ln\left(\frac{1}{\widetilde{h}}\right)+\frac{1}{3}\left(\frac{\gamma_{E}t}{\tilde{h}}\right)^{2}\right].
\end{split}
\end{equation*}
Performing the analytical continuation $\omega_{\nu}\rightarrow-i\omega$, one finds the corresponding ac-conductivity
\begin{equation*}
\sigma_{xx}^{\mathrm{DOS}(9-10)}(\omega)=i\frac{e^{2}}{\pi^{2}}\frac{\Delta}{\omega}\left\{\!\left(\frac{\Delta}{\omega}\right)^{2}\!\ln\left[\frac{\left(1+\frac{i\omega}{\Delta}\right)\left(1-\frac{2i\omega}{\Delta}\right)}{1-\frac{i\omega}{\Delta}}\right]\!+\!\frac{2}{1-\frac{2i\omega}{\Delta}}\right\}\! \left\{\!\ln\left(\frac{1}{\widetilde{h}}\right)\!+\!\frac{1}{3}\left(\frac{\gamma_{E}t}{\tilde{h}}\right)^{2}\!\left[2-\frac{1}{\left(1-\frac{i\omega}{\Delta\widetilde{h}}\right)^{2}}\right]\!+\!\ln\left(1-\frac{i\omega}{\Delta\widetilde{h}}\right)\!\right\} . 
\end{equation*}
The low frequency asymptotics reads
\begin{align*}
\sigma_{xx}^{\mathrm{DOS}(9-10)}(\omega\ll\Delta\widetilde{h})=\frac{2e^{2}}{\pi^{2}}\left[-\frac{i\omega}{\Delta\widetilde{h}}+\frac{1}{2}\left(\frac{\omega}{\Delta\widetilde{h}}\right)^{2}+\ln\left(\frac{1}{\widetilde{h}}\right)+\frac{1}{3}\left(\frac{\gamma_{E}t}{\tilde{h}}\right)^{2}\right],
\end{align*}
which cancels exactly the DOS(5-8) contribution in Eq. (\ref{sgima58}).


\begin{thebibliography}{99}
\bibitem{footnote}
One should remember that the fluctuation correction calculated in the first order of perturbation theory must remain smaller than the corresponding value of the Drude conductivity. This requirement restricts the range of magnetic fields in which our theory is valid to
$\frac{\hbar^{2}}{p_F^2 l d} \lesssim \tilde{h} \ll 1,$
where $d$ is the film thickness. The lower bound of this restriction coincides with the Ginzburg–Levanyuk number for a $2D$ superconducting film. It is worth noting that its value in the quantum fluctuation regime turns out to be the same as in the vicinity of the critical temperature \cite{LV09}.
\end{thebibliography}
\end{document}